\documentclass[twocolumn,aps,prd,eqsecnum,nofootinbib]{revtex4}
\usepackage{verbatim}
\usepackage{amssymb}
\usepackage{amsmath}
\usepackage{amsfonts}
\usepackage{bm}

\begin{document}

\newcommand{\nin}{\noindent}
\newcommand{\nnm}{\nonumber}
\newcommand{\doe}{\partial}
\newcommand{\be}{\begin{equation}}
\newcommand{\ee}{\end{equation}}
\newcommand{\bea}{\begin{eqnarray}}
\newcommand{\eea}{\end{eqnarray}}
\newcommand{\bdm}{\begin{displaymath}}
\newcommand{\edm}{\end{displaymath}}
\newcommand{\bse}{\begin{subequations}}
\newcommand{\ese}{\end{subequations}}
\newcommand{\tb}{\textbf}
\newcommand{\n}{\,^{\scriptscriptstyle \text{n}}\!\;\!}
\newcommand{\pn}{\,^{\scriptscriptstyle \text{pn}}\!\;\!}
\newcommand{\cm}{\,^{\scriptscriptstyle \text{cm}}\!\;\!}
\newcommand{\bu}{\nin$\bullet\:\:$}
\newcommand{\mc}{\mathcal}
\newcommand{\gA}{{{\rm{g}},A}}
\newcommand{\gB}{{{\rm{g}},B}}
\newcommand{\go}{{{\rm{g}},1}}
\newcommand{\gt}{{{\rm{g}},2}}
\newcommand{\g}{{\rm{g}}}
\newcommand{\sys}{{\rm{sys}}}

\title{First-post-Newtonian quadrupole tidal
interactions in binary systems}

\author{Justin E. Vines}
\author{\'Eanna \'E. Flanagan}
\affiliation{Department of Physics, Cornell University, Ithaca, NY 14853, USA.}
\date{\today}

\begin{abstract}

We consider tidal coupling in a binary stellar system to first-post-Newtonian order.  We derive the orbital equations of motion for bodies with spins and mass quadrupole moments and show that they conserve the total linear momentum of the binary.  We note that spin-orbit coupling must be included in a 1PN treatment of tidal interactions in order to maintain consistency (except in the special case of adiabatically induced quadrupoles); inclusion of 1PN quadrupolar tidal effects while omitting spin effects would lead to a failure of momentum conservation for generic evolution of the quadrupoles.  We use momentum conservation to specialize our analysis to the system's center-of-mass-energy frame; we find the binary's relative equation of motion in this frame and also present a generalized Lagrangian from which it can be derived.  We then specialize to the case in which the quadrupole moment is adiabatically induced by the tidal field (in which case it is consistent to ignore spin effects).  We show how the adiabatic dynamics for the quadrupole can be incorporated into our action principle and present the simplified orbital equations of motion and conserved energy for the adiabatic case.  These results are relevant to gravitational wave signals from inspiralling binary neutron stars.

\end{abstract}

\maketitle

\section{Introduction and summary}\label{intro}

\subsection{Background and motivation}\label{context}

Inspiralling and coalescing compact binaries present one of the most promising sources for ground-based gravitational wave (GW) detectors \cite{CutlerThorne}.  A primary goal in the measurement of GW signals from neutron star-neutron star (NSNS) and black hole-neutron star (BHNS) binaries is to probe the neutron star matter equation of state (EoS), which is currently only loosely constrained by electromagnetic observations in the relevant density range $\rho\sim$ 2--8$\times 10^{14}$ g/cm$^3$ \cite{EOS}.   The EoS will leave its imprint on the GW signal via the effects of tidal coupling, as the neutron star is distorted by the non-uniform field of its companion.  Many recent studies of the effects of the neutron star EoS on binary GW signals have been based on numerical simulations of the fully relativistic hydrodynamical evolution of NSNS and BHNS binaries (see e.g.~the reviews \cite{Duez,Faber}).  These simulations have largely focused on the binaries' last few orbits and merger, at GW frequencies $\gtrsim$500 Hz, and have investigated constraining neutron star structure via (for example) the GW energy spectrum \cite{spectrum}, effective cutoff frequencies at merger \cite{cutoff}, and tidal disruption signals \cite{disrupt}.

As recently investigated by Flanagan and Hinderer \cite{FH}, neutron star internal structure may also have a measurable influence on the GW signal from the earlier inspiral stage of a binary's orbital evolution, at GW frequencies $\lesssim$500 Hz.  While tidal coupling will produce only a small perturbation to the GW signal in this low-frequency adiabatic regime, the tidal signal should be relatively clean, depending (at leading order) on a single parameter pertaining to the neutron star structure.  This tidal deformability parameter $\lambda$ is the proportionality constant between the applied tidal field and the star's induced quadrupole moment and is sensitive to the neutron star EoS.  The measurement scheme proposed in Ref.~\cite{FH} is based on an analytical model for the tidal contribution to the GW signal, giving a linear perturbation to the GW phase proportional to $\lambda$.  At GW frequencies $\lesssim$400 Hz, the model should be sufficiently accurate to constrain $\lambda$ to $\sim$10\%, with the largest source of error being first-post-Newtonian (1PN) corrections to the tidal-orbital coupling \cite{FH,Hetal}.  Recent work in Ref.~\cite{villain} suggests that the the effective one body (EOB) formalism (discussed further below) can be used to extend the description of the GW phasing up to merger, allowing Advanced LIGO to detect and measure tidal polarizations of neutron stars.

The modeling of tidal effects in GW signals from neutron star binaries can be divided into three separate problems:  (i) to calculate the deformation response of each neutron star to the tidal field generated by its companion, (ii) to calculate the influence of the tidal deformations on the system's (conservative) orbital dynamics, and (iii) to calculate the gravitational waveform emitted by the system and incorporate the corresponding radiation reaction effects in the orbital dynamics.  While this paper is focused on solving problem (ii) to 1PN accuracy, we will briefly mention work on each of these problems.

\emph{Tidal Deformability Computations:}  While computing tidal deformations of stars is a well studied problem in Newtonian gravity \cite{Love}, it has recently been re-examined in the context of fully relativistic stars by Hinderer et al.~\cite{H,Hetal}, Damour and Nagar \cite{DN1}, and Binnington and Poisson \cite{BP}.  These authors used fully relativistic models for the star's interior, with various candidate equations of state, to calculate the perturbations to the star's equilibrium configuration produced by a static external tidal field.  In Refs.~\cite{H,Hetal}, the results were used to determine the (electric-type) quadrupolar tidal deformability $\lambda$, while Refs.~\cite{DN1,BP} corrected computational errors in Ref.~\cite{Hetal} and extended the analysis and included all higher-multipolar electric-type response coefficients (for the mass multipole moments) as well as their magnetic-type analogues (for the current multipole moments).  While these results concern a star's response to a static tidal field, they are still applicable for inspiralling binaries at sufficiently low orbital frequencies; a star will approximately maintain equilibrium with the instantaneous tidal field if the field changes adiabatically.  

\emph{Orbital-tidal conservative dynamics:} In their treatment of tidal effects in binary GW signals in Ref.~\cite{FH}, Flanagan and Hinderer used Newtonian gravity to treat (the conservative part of) the system's orbital dynamics; they estimated that 1PN corrections to the orbits would modify the calculated tidal signal by $\sim$10\%.   A general formalism for calculating 1PN corrections to the orbital dynamics of extended bodies has been developed in a series of papers \cite{DSX1,DSX2,DSX3,DSX4} by Damour, Soffel, and Xu (DSX) and later extended by Racine and Flanagan \cite{RF}.  In this paper, we apply that formalism to determine and analyze the explicit 1PN translational equations of motion for binary systems with quadrupolar tidal interactions.  While such equations of motion have previously been presented in Refs.~\cite{XuWuSchafer,WuHeXu}, we find that those results differ from ours.  We also extend previous results by analyzing 1PN momentum conservation for the binary system (which serves as a strong consistency check for our equations of motion, and one not satisfied by the results of Refs.~\cite{XuWuSchafer,WuHeXu}), by specializing the equations of motion to the system's center-of-mass-energy frame, and by formulating an action principle for the orbital dynamics.

Tidal effects in the conservative dynamics to 1PN order have also been investigated recently by Damour and Nagar (DN) \cite{DN2} who considered circular orbits, and (since the first appearance of this work) by Bini, Damour and Faye (BDF) \cite{BDF}, who considered generic orbits and have in fact carried the analysis to 2PN order.  These works incorporate their results into an EOB description of the dynamics.  In Appendix \ref{dam}, we demonstrate that our results agree with those of DN and with (the 1PN restriction of) those of BDF.

\emph{Gravitational Wave Emission:} To calculate the emitted GW signal and radiation reaction effects, Flanagan and Hinderer \cite{FH} used the quadrupole radiation approximation and associated 2.5PN radiation reaction forces.  To consistently generalize their calculation in Ref.~\cite{FH} to (relative) 1PN order, it is necessary to compute the 3.5PN corrections to the GW generation, in addition to the 1PN orbital corrections considered in this paper.  These calculations have been carried out by Vines, Hinderer and Flanagan in Ref.~\cite{VHF}, which builds off of the work presented here.  The issue of tidal effects in inspiral-stage NSNS binary GW signals has also been recently addressed by means of numerical relativity simulations in Refs.~\cite{Read,Baiotti}.

We now turn to a more detailed overview of the problem of the conservative orbital-tidal dynamics, at Newtonian order in Sec.~\ref{intronewton} and at 1PN order in Sec.~\ref{intropn}, before summarizing the results of this paper in Sec.~\ref{results}.  

\subsection{Newtonian tidal coupling}\label{intronewton}

Gravitational tidal coupling arises from the interaction of the non-spherical components of a body's matter distribution with a non-uniform gravitational field.  The non-sphericity is characterized at leading order by the body's mass quadrupole moment,
\be\label{introQ}
Q^{ij}(t)=\int d^3x \,\rho\,(\bar x^i\bar x^j-\frac{1}{3}\delta^{ij}|\bar{\bm x}|^2),
\ee
with $\rho(t,\bm x)$ being the mass density and $\bar{x}^i(t)=x^i-z^i(t)$ being the displacement from the body's center of mass position $x^i=z^i(t)$.  The quadrupole is (at most) on the order of $Q^{ij}\sim MR^2$, with $M$ being the body's mass and $R$ its radius.  Higher-order deformations are described by the octupole, $Q^{ijk}\sim\int d^3x \,\rho\,\bar x^i\bar x^j\bar x^k\sim MR^3$, and higher-order multipole moments.  The non-uniform field can be characterized by derivatives of an external Newtonian potential $\phi_{\rm{ext}}(t,\bm{x})$; in a binary system, the leading-order potential is $\phi_{\rm{ext}}=-GM/r$, where $G$ is Newton's constant, $M$ is the mass of the companion, and $r$ the distance between the body and its companion.

The non-uniform field of the companion produces tidal forces on the non-spherical body (in addition to the usual $1/r^2$ force) according to
\bea
M\ddot z^i&=&-\left(M\doe_i+\frac{1}{2}Q^{jk}\doe_{ijk}+\frac{1}{6}Q^{jkl}\doe_{ijkl}+\ldots\right)\phi_{\rm{ext}}
\nnm\\
&\sim&GM\left[\frac{M}{r^2}+\frac{|Q^{ij}|}{r^4}+\frac{|Q^{ijk}|}{r^5}+\ldots\right]
\nnm\\ \label{schematic}
&\sim& \frac{GM^2}{r^2}\left[1+\frac{R^2}{r^2}+O\left(\frac{R^3}{r^3}\right)\right],
\eea
where $M$ is the mass of the body or its companion, assumed here to be roughly equal, and the derivatives of the external potential are evaluated at the body's center of mass, $x^i=z^i(t)$.  The contributions to the net force from the quadrupole and higher-order multipoles are thus seen to take the form of an expansion in $R/r$, the ratio of the size of the body to the orbital separation.  When this finite-size parameter is small, as in the early stages of binary inspiral, the tidal force is well approximated by the quadrupolar term alone.  A more detailed account of Newtonian tidal forces and related results is given in Sec.~\ref{newton} below.

In neutron star binaries, a quadrupole is induced by differential forces resulting from the non-uniform field of the companion.  In the adiabatic limit, when the response time scale of the body is much less than the time scale on which the tidal field changes, the induced quadrupole will be given (to linear order in the tidal field) by
\be\label{introlambda}
Q^{ij}(t)=-\lambda\doe_i\doe_j\phi_{\rm{ext}}(t,\bm{z}),
\ee
with the constant $\lambda$ being the tidal deformability.  This is related to the more often used dimensionless Love number $k_2$ by $\lambda=2k_2R^5/3G$, where $R$ is the body's radius \cite{Love}.

Using Newtonian gravity to describe the orbital dynamics and the adiabatic approximation to model the stars' induced quadrupoles, Flanagan and Hinderer \cite{FH} calculated the effect of tidal interactions on the phase of the gravitational waveform emitted by an inspiralling neutron star binary and analyzed the measurability of the tidal effects.  They found that Advanced LIGO should be able to constrain the neutron stars' tidal deformability to $\lambda\le (2.0\times 10^{37}\,\rm{g}\,\rm{cm}^2\,\rm{s}^2)(D/50\,\rm{Mpc})$ with 90\% confidence, for a binary of two 1.4 $M_\odot$ neutron stars at a distance $D$ from the detector, using only the portion of the signal with GW frequencies less than 400 Hz.  The calculations of $\lambda$ for a 1.4 $M_\odot$ neutron star in Refs.~\cite{H,Hetal,DN1,BP}, using several different equations of state, give values in the range 0.03--1.0$\times 10^{37}\,\rm{g}\,\rm{cm}^2\,\rm{s}^2$, so that nearby events may allow Advanced LIGO to place useful constraints on candidate equations of state.  Reference \cite{villain} discusses how the EOB formalism can be used to extend the range of validity of analytic waveforms up to merger, and argues that Advanced LIGO should be able to detect and measure the tidal deformability of neutron stars.

Refs.~\cite{FH,Hetal} estimate the fractional corrections to the tidal signal due to several effects neglected by the model of the GW phasing used in Ref.~\cite{FH}, namely, non-adiabaticity ($\lesssim$1\%), higher-multipolar tidal coupling ($\lesssim$0.7\%), nonlinear hydrodynamic effects ($\lesssim$0.1\%), spin effects ($\lesssim$0.3\%), nonlinear response to the tidal field ($\lesssim$3\%), viscous dissipation (negligible), and post-Newtonian effects ($\lesssim$10\%).  The largest corrections, from post-Newtonian effects in the orbital dynamics and GW emission, will depend on the neutron star physics only through the same tidal deformability parameter $\lambda$ used in the Newtonian treatment and thus can be easily incorporated into the data analysis methods outlined in Refs.~\cite{FH,Hetal}.  

\subsection{First-post-Newtonian corrections}\label{intropn}

For inspiralling neutron star binaries with a total mass of $\sim$$3 M_\odot$ at orbital frequencies of $\sim$200 Hz (GW frequencies of $\sim$400 Hz), the post-Newtonian expansion parameter $v^2/c^2\sim GM/c^2r$ is $\sim$0.1, so the 1PN approximation is well suited to describing relativistic corrections to the binary orbit.  As discussed in depth in Sec.~\ref{pnti} below, the 1PN orbital dynamics of a binary system with tidal coupling can be described by translational equations of motion (EoMs) similar in form to the Newtonian equations schematically represented in Eq.~(\ref{schematic}), giving the center-of-mass acceleration of each constituent body in terms of their positions and multipole moments.  The 1PN equations of motion add order $1/c^2$ correction terms to (\ref{schematic}) which depend not only on the bodies' positions and mass multipole moments, $M$, $Q^{ij}$, $Q^{ijk}$, etc., but also on their velocities and current multipole moments, $S^i$, $S^{ij}$, etc.  Expanding in both the post-Newtonian parameter $v^2/c^2$ and the finite size parameter $R/r$, the 1PN equations of motion can be written schematically as
\bea
M\ddot{z}^i&\sim&GM\bigg[\frac{M}{r^2}+\frac{v^2}{c^2}\frac{M}{r^2}+O\left(\frac{v^4}{c^4}\frac{M}{r^2}\right)
\nnm\\
&&+\frac{|Q^{ij}|}{r^4}+\frac{v^2}{c^2}\frac{|Q^{ij}|}{r^4}+O\left(\frac{|Q^{ijk}|}{r^5}\sim\frac{R^3}{r^3}\frac{M}{r^2}\right)
\nnm\\ \label{pnschematic}
&&+\frac{v|S^i|}{c^2r^3}+O\left(\frac{v|S^{ij}|}{c^2r^4}\sim\frac{v^2}{c^2}\frac{R^2}{r^2}\frac{M}{r^2}\right)\bigg]
\eea
In the top line, the first term gives the usual point-particle (monopole) force of Newtonian gravity, and the second term represents its 1PN corrections.  The last term of the top line denotes 2PN and higher post-Newtonian order corrections, which we neglect in this paper.  Point-particle EoMs are in fact currently known up through 3PN order \cite{3p5}.

In the second line of Eq.~(\ref{pnschematic}), we have first the Newtonian quadrupole-tidal term, followed by its 1PN corrections (which are the subject of this paper), and finally contributions from the octupole and higher mass multipoles (and their post-Newtonian corrections) which are suppressed by higher powers of $R/r$ and which we neglect in our analysis \cite{RF}.

The first term in the bottom line represents the 1PN spin-orbit coupling \cite{DSX2}, which will be included in our analysis, and the final term of the bottom line denotes 1PN contributions from the bodies' current quadrupoles and higher current multipoles, which will not be included.

The DSX treatment of 1PN celestial mechanics \cite{DSX1,DSX2,DSX3,DSX4} provides a framework for calculating the orbital EoMs for bodies with arbitrarily high-order mass and current multipole moments.  In Ref.~\cite{DSX2}, DSX applied their formalism to rederive the explicit 1PN EoMs for bodies with mass monopoles and current dipoles (spins).  The calculation was then extended to include the effects of the bodies' mass quadrupoles by Xu et al \cite{XuWuSchafer,WuHeXu}.  Racine and Flanagan (RF) \cite{RF} later reworked the DSX formalism and presented explicit 1PN EoMs for bodies with arbitrarily high-order multipoles, which can be specialized to the case of bodies with only spins and mass quadrupoles.  We have found, however, that the final results of Xu et al and those of RF are in disagreement with each other and with our recent calculations.  Some typos and omissions leading to errors in RF have been identified and are outlined in an upcoming erratum; the results given in this paper agree with the corrected results of RF.  In Sec.~\ref{pnti} below, we review the essential ideas of the DSX formalism, following the notations and conventions of RF, and we outline the full procedure by which our results for the 1PN EoMs are derived.

Though it would be convenient to be able to specialize to the case of non-spinning bodies when studying tidal interactions, this would lead to inconsistencies at 1PN order when considering generic behavior of the quadrupole moments.  A body with a quadrupole in a tidal field will generically experience tidal torques [according to Eq.~(\ref{introSdot}) below] which will spin up the body even if it started with no spin; this is a Newtonian-order effect.  The resultant spin affects the orbital dynamics via the 1PN spin-orbit coupling.  For this reason, if one were to work through the DSX formalism and simply drop all spin terms while keeping mass quadrupole terms, one would arrive at inconsistencies.  In particular, one would find that momentum is not conserved at 1PN order (see Sec.~\ref{sysdipole} below).  However, for the special case of adiabatically induced quadrupoles, the tidal torques vanish, and it is then consistent to ignore all spin terms.

The relation between the work of DN \cite{DN2} and BDF \cite{BDF} on the 1PN conservative orbital-tidal dynamics and the results of this paper is described in the next subsection along with the summary of our results, and in more detail in Appendix \ref{dam}.  We also refer the reader to Refs.~\cite{villain,baiotti3,vega,Read2,tsang,ferraritwo,messenger,bernuzzi2,maselli,bernuzzi} for other recent work on tidal effects in inspiralling binaries.

\subsection{Summary of results}\label{results}

\subsubsection{The $M_1$-$M_2$-$S_2$-$Q_2$ system}

Our results concern the 1PN gravitational interactions in a system of two bodies, which we label ``1'' and ``2''.  We model body 1 as an effective point particle, with a mass monopole moment only, while we take body 2 to have additionally a spin and a mass quadrupole moment.  We consistently work to linear order in the spin and quadrupole, and our results can thus be easily generalized to the case of two spinning, deformable bodies by interchanging particle labels.  We initially assume nothing about the bodies' internal structure or dynamics.  Our primary assumption is the validity of the 1PN approximation to general relativity in a vacuum region surrounding the bodies.

The system's orbital dynamics can be formulated in terms of the bodies' center-of-mass worldlines $z_1^i(t)$ and $z_2^i(t)$ and their multipole moments: the mass monopoles $M_1(t)$ and $M_2(t)$, the spin $S_2^i(t)$, and the mass quadrupole $Q_2^{ij}(t)$.

The worldlines $x^i=z_1^i(t)$ and $x^i=z_2^i(t)$ parametrize the bodies' positions to 1PN accuracy in a (conformally Cartesian and harmonic) 'global' coordinate system $(t,x^i)$, which tends to an inertial coordinate system in Minkowski spacetime as $|\bm x|\to\infty$.  The global coordinates and center-of-mass worldlines are defined more precisely in Sec.~\ref{global}.  We use the following notation for the relative position and velocity:
\bea
&z^i=z_2^i-z_1^i,\quad r=|\bm{z}|=\sqrt{\delta_{ij}z^iz^j},\quad n^i=z^i/r,& \nnm
\\ \nnm
&v_1^i=\dot z_1^i,\quad v_2^i=\dot z_2^i,\quad v^i=v_2^i-v_1^i,\quad \dot r=v^in^i,
\eea
with dots denoting derivatives with respect to the global frame time coordinate $t$.

The multipole moments---$M_1(t)$, $M_2(t)$, $S_2^i(t)$, and $Q_2^{ij}(t)$ for our truncated system---are defined in Secs.~\ref{body} and \ref{loms} via a multipole expansion of the 1PN metric in a vacuum region surrounding each body \cite{RF}.  In the case of weakly self-gravitating bodies, these moments can be defined as integrals of the stress-energy tensor over the volume of the bodies, as in Eqs.~(\ref{stressint}); these definitions coincide with those of the Blanchet-Damour multipole moments introduced in Ref.~\cite{BD} and used by DSX \cite{DSX1,DSX2,DSX3,DSX4}.  The mass multipole moments (like $M_1$, $M_2$, and $Q_2^{ij}$) are defined with 1PN accuracy, while the current multipole moments (like $S_2^i$) appear only in 1PN-order terms and thus need only be defined with Newtonian accuracy.  We will often denote the spin and quadrupole of body 2 by $S_i$ and $Q^{ij}$, dropping the ``2'' labels.

\subsubsection{General equations of motion and orbital Lagrangian}

The equations of motion for the monopoles $M_1(t)$ and $M_2(t)$, the spin $S^i(t)$, and the worldlines $z_1^i(t)$ and $z_2^i(t)$ are determined by Einstein's equations alone, while that for the quadrupole $Q^{ij}(t)$ will depend on the details of body 2's internal dynamics and can initially be left unspecified.  The mass monopole of body 1, the effective point particle, is found to be conserved to 1PN order:
\be
\dot M_1=O(c^{-4}),
\ee
while that of body 2 is not.  As discussed in Sec.~\ref{truncate}, the 1PN-accurate mass monopole $M_2$ can be decomposed according to
\bse\label{introM2dot}
\be\label{introM2}
M_2=\n M_2+c^{-2}\left(E_2^{\rm{int}}+3U_Q\right)+O(c^{-4}).
\ee
Here, $\n M_2$ is the Newtonian-order (rest mass) contribution, which is conserved:
\be\label{intronM2}
\n\dot M_2=0.
\ee
The 1PN contributions to (\ref{introM2}) involve the Newtonian potential energy of the quadrupole-tidal interaction,
\be\label{introUQ}
U_Q=-\frac{3M_1}{2r^3}Q^{ij}n^in^j,
\ee
and the Newtonian internal energy of body 2, $E_2^{\rm{int}}$, whose evolution is governed by the rate at which the tidal field does work on the body \cite{Thorne}:
\be\label{introE2intdot}
\dot E_2^{\rm{int}}=\frac{3M_A}{2r^3}\dot Q^{ij}n^in^j,
\ee
\ese
(cf.~Sec.~\ref{nEnergy}).  Equations (\ref{introM2dot}) ensure that $M_2$ satisfies the 1PN evolution equation (\ref{M2dot},\ref{nG2ij}).  The decomposition of the monopole $M_2$ in (\ref{introM2}) is essential for properly formulating an action principle for the orbital dynamics.  The evolution of the spin $S^{i}$ is determined by the (Newtonian-order) tidal torque formula:
\be\label{introSdot}
\dot S^i=\frac{3M_1}{r^3} \epsilon^{ijk} Q^{ja}n^an^k+O(c^{-2}),
\ee
as in (\ref{nspineom}).  Finally, the 1PN translational equations of motion, which govern the evolution of the worldlines $z_1^i$ and $z_2^i$, are of the form
\bea
\ddot z_1^i&=&\mc F_1^i(z^j,v_1^j,v_2^j,M_1,M_2,S^j,Q^{jk},\dot Q^{jk},\ddot Q^{jk}),
\nnm\\ 
\ddot z_2^i&=&\mc F_2^i(z^j,v_1^j,v_2^j,M_1,M_2,S^j,Q^{jk},\dot Q^{jk}), \nnm
\eea
and are given explicitly by Eqs.~(\ref{teoms}).

In Sec.~\ref{system}, we define and calculate the 1PN-accurate mass dipole moment of the entire system $M^i_{\rm{sys}}(t)$, given in Eq.~(\ref{Misys}).  We find that the condition $\ddot M^i_{\rm{sys}}=O(c^{-4})$, required by Einstein's equations and reflecting the conservation of the system's total momentum, is satisfied as a consequence of the orbital EoMs (\ref{teoms}); this serves as a non-trivial consistency check for our results.  The conservation of momentum also allows us to specialize the EoMs to the system's center-of-mass(-energy) (CoM) frame, which can be defined by the condition $M^i_{\rm{sys}}=0$ as in Sec.~\ref{comframe}.  The two EoMs (\ref{teoms}) for the worldlines $z_1^i$ and $z_2^i$ can then be traded for the single EoM for the CoM-frame relative position $z^i=z_2^i-z_1^i$, given in Eq.~(\ref{aCoM}).

Our results can be most compactly summarized by giving a Lagrangian formulation of the CoM-frame orbital dynamics, as discussed in Sec.~(\ref{lagrangian}).  We find that the CoM-frame orbital EoM (\ref{aCoM}) can be derived from the generalized Euler-Lagrange equation
\be\label{introEL}
\left(\frac{\doe}{\doe z^i}-\frac{d}{dt}\frac{\doe}{\doe v^i}+\frac{d^2}{dt^2}\frac{\doe}{\doe a^i}\right)\mc L_{\rm{orb}}=0,
\ee
with a generalized Lagrangian $\mc L_{\rm{orb}}$ given by
\bse\label{introL}
\be\label{introLsplit}
\mc L_{\rm{orb}}=\mc L_M+\mc L_S+\mc L_Q,
\ee
with the monopole part,
\bea\label{introLM}
\mc L_M&=&\frac{\mu v^2}{2} + \frac{\mu M}{r} 
+\frac{\mu}{c^2}\bigg[ \frac{1-3\eta}{8}v^4 
\\ \nnm
&&+ \frac{M}{2r} \left( \eta \dot{r}^2 + (3+\eta)v^2 - \frac{M}{r} \right)\bigg]+O(c^{-4}),
\eea
the spin part,
\be\label{introLS}
\mc L_S=\frac{\chi_1}{c^2}\epsilon^{abc}S^av^b\left[\frac{2M}{r^2}n^c+\frac{\chi_1 }{2}a^c\right]+O(c^{-4}),
\ee
and the quadrupole part,
\begin{widetext}
\bea\label{introLQ}
\mc L_Q&=&\frac{3M_1}{2 r^3} Q^{ab} n^an^b 
+ \frac{1}{c^2} \Bigg\{ \frac{M}{r^3}Q^{ab} \left[ n^an^b \left( A_1 v^2 + A_2 \dot{r}^2 + A_3 \frac{M}{r} \right) + A_4 v^av^b + A_5 \dot{r} n^a v^b \right]
\nnm\\
&& \phantom{\frac{3M_1}{2 r^3} Q^{ab} n^an^b + \frac{1}{c^2} \Bigg\{} + \frac{M}{r^2}\dot{Q}^{ab}\left[A_6 n^a v^b + A_7 \dot{r} n^an^b \right] 
 + E^{\rm{int}}_2\left[A_8 v^2 + A_9 \frac{M}{r} \right] \Bigg\}+O(c^{-4}).
\eea
\end{widetext}
\ese
We use here the notation
\bea
&M=M_1+\n M_2,\quad \chi_1=M_1/M,\quad \chi_2=\n M_2/M,&
\nnm\\ \nnm
&\mu=M_1 {\n M_2}/M,\quad \eta=\chi_1\chi_2=\mu/M,&
\eea
with $M$ being the total (conserved) Newtonian rest mass, $\mu$ the reduced mass, and $\eta$ the symmetric mass ratio.  The dimensionless coefficients $A_1$--$A_9$ appearing in (\ref{introLQ}) are functions only of the mass ratios $\chi_1$ and $\chi_2$ given by (\ref{As}).  

\subsubsection{Adiabatic approximation for the induced quadrupole}

The above results concerning the orbital EoM and its Lagrangian formulation are valid regardless of the internal structure of body 2, i.e.~for arbitrary evolution of its quadrupole $Q^{ij}(t)$.  In Sec.~\ref{adiabatic}, we discuss a simple adiabatic model for the evolution of $Q^{ij}$.  In the adiabatic limit, the (body-frame) quadrupole responds to the instantaneous tidal field according to
\be\label{intropnlambda}
Q^{ij}(t)=\lambda G_2^{ij}(t),
\ee
where $G_2^{ij}(t)$, given by Eq.~(\ref{pnG2}), is the (body-frame) quadrupolar gravito-electric tidal moment of body 2, a 1PN generalization of the derivatives of the Newtonian potential in Eq.~(\ref{introlambda}), and $\lambda$ is the (constant) tidal deformability.  With the quadrupole given by (\ref{intropnlambda}), the tidal torque (\ref{introSdot}) vanishes (cf. Eq.~(\ref{nadspin})), so that the spin $S^i$ is constant; thus, in the adiabatic limit, we can specialize to the case $S^i=0$ without generating inconsistencies.  In Sec.~\ref{ELQ}, we show that the adiabatic evolution for the quadrupole (\ref{intropnlambda}) can be derived from a Lagrangian that adds to the orbital Lagrangian an internal elastic potential energy term which is quadratic in the 1PN-accurate quadrupole:
\be\label{introLad}
\mc L=\mc L_{\rm{orb}}[z^i,Q^{ij}]-\frac{1}{4\lambda}Q^{ab}Q^{ab}+O(c^{-4}),
\ee
with $\mc L_{\rm{orb}}$ given by Eq.~(\ref{introL}) with $S^i=0$, and with $E^{\rm{int}}_2=(1/4\lambda)Q^{ab}Q^{ab}+O(c^{-2})$.  (Note that any additional constant contribution to the internal energy $E^{\rm{int}}_2$ can be included as a 1PN contribution to the constant $\n M_2$ in Eq.~\ref{introM2}.)  Substituting the solution (\ref{intropnlambda}) for $Q^{ij}(t)$ into this Lagrangian, we obtain a simplified Lagrangian involving only the CoM-frame relative position $z^i(t)$:
\begin{widetext}
\be\label{introLr}
\mc L[z^i]=\frac{\mu v^2}{2}+\frac{\mu M}{r}\left(1+\frac{\Lambda}{r^5}\right)
+\frac{\mu}{c^2}\left\{\theta_0v^4
+\frac{M}{r}\left[v^2\left(\theta_1+\xi_1\frac{\Lambda}{r^5}\right)
+\dot{r}^2\left(\theta_2+\xi_2\frac{\Lambda}{r^5}\right)
+\frac{M}{r}\left(\theta_3+\xi_3\frac{\Lambda}{r^5}\right)\right]\right\}
\ee
\end{widetext}
with $\Lambda=(3\chi_1/2\chi_2)\lambda$, and with the dimensionless coefficients
\bea
\theta_0&=&(1-3\eta)/8,
\nnm\\
\theta_1&=&(3+\eta)/2,
\nnm\\
\theta_2&=&\eta/2,
\nnm\\
\theta_3&=&-1/2,
\nnm\\
\xi_1&=&(\chi_1/2)(5+\chi_2),
\nnm\\ 
\xi_2&=&-3(1-6\chi_2+\chi_2^2),
\nnm\\
\xi_3&=&-7+5\chi_2.
\eea

This Lagrangian represents one of the primary results of this paper.  Since the first appearance of this work, an analogous Lagrangian (in a different gauge) has been derived by Bini, Damour and Faye in Ref.~\cite{BDF}, using effective action techniques.  In fact, BDF have greatly extended the analysis by carrying the calculation to 2PN order.  Also (prior to this work), Damour and Nagar \cite{DN2} presented a 1PN Hamiltonian valid for circular orbits.  These works both incorporate their results into the EOB formalism.  In Appendix \ref{dam}, we demonstrate the complete equivalence of those results with ours at 1PN order.

The orbital EoM resulting from the Lagrangian (\ref{introLr}), which can also be found by substituting the adiabatic solution (\ref{intropnlambda}) for $Q^{ij}(t)$ directly into the general orbital EoM (\ref{aCoM}), is given by Eq.~(\ref{ada}).  The conserved energy $E(\bm{z},\bm{v})$ derived from the Lagrangian (\ref{introLr}) is given by Eq.~(\ref{Ead}).  In the case of circular orbits, we find the gauge-invariant energy-frequency relationship
\begin{widetext}
\be
E(\omega)=\mu(M\omega)^{2/3}\left[-\frac{1}{2}+\frac{9\chi_1}{2\chi_2}\frac{\lambda\omega^{10/3}}{M^{5/3}}
+\frac{(9+\eta)}{24}\frac{(M\omega)^{2/3}}{c^2}+\frac{11\chi_1}{4\chi_2}(3+2\chi_2+3\chi_2^2)\frac{\lambda\omega^4}{Mc^2}\right].
\ee
\end{widetext}
This result, along with others from this paper, are used in calculating the phasing of GW signals from inspiralling neutron star binaries in Ref.~\cite{VHF}.

\subsection{Notation and Conventions}\label{notation}

We use units where Newton's constant is $G=1$, but retain factors of the speed of light $c$, with $1/c^2$ serving as the formal expansion parameter for the post-Newtonian expansion.  We use lowercase Latin letters $a,b,i,j,\ldots$ for indices of (three-dimensional) spatial tensors.  Spatial indices are contracted with the Euclidean metric, $v^iw^i=\delta_{ij}v^iw^j$, with up or down placement of the indices having no meaning.  We use uppercase Latin letters to denote multi-indices: $L$ denotes the $l$ indices $a_1a_2\ldots a_l$, $K$ denotes the $k$ indices $b_1b_2\ldots b_k$, etc.  For a given vector $v^i$ or for the partial derivative operator $\doe_i$, we use multi-indices or explicit sequences of indices to denote their tensorial powers:
\bea
v^L&=&v^{a_1a_2\ldots a_l}=v^{a_1}v^{a_2}\ldots v^{a_l}, 
 \nnm \\
\doe_K&=&\doe_{b_1b_2\ldots b_k}=\doe_{b_1}\doe_{b_2}\ldots\doe_{b_k}, \label{tensorpower}
\eea
and, for example, $v^{ij}=v^iv^j$.  We also use $v^2=v^{ii}$ and $\nabla^2=\doe_{ii}$.  Multi-indices are also used for sets of distinct tensors of varying rank, $\{M,M^a,M^{ab},\ldots\}$, with $M^L=M^{a_1a_2\cdots a_l}$ denoting the tensor of rank $l$.  We use the Einstein summation convention for both individual indices and multi-indices.  Derivatives with respect to a time coordinate $t$ are denoted by $\doe_t$ or by overdots.

We use angular brackets to denote the symmetric, trace-free (STF) projection of tensors \cite{DSX1}:
\bea
T^{<ab>}&=&T^{(ab)}-\frac{1}{3}\delta^{ab}T^{cc},
\nnm\\
T^{<abc>}&=&T^{(abc)}-\frac{3}{5}\delta^{(ab}T^{c)dd},  \label{STF3}
\eea
and so on, with parentheses denoting the symmetric projection.  For a STF tensor $S^L=S^{<L>}$ and general tensor $T^L$, note that 
\be\label{STFcontract}
S^LT^L=S^LT^{<L>}.
\ee

\section{Newtonian tidal interactions}\label{newton}

In this section, we review the standard treatment of tidal coupling in Newtonian theory \cite{Love}; the first-post-Newtonian treatment given subsequent sections makes extensive use of these Newtonian-order results.  We define the multipole moments and tidal moments of an extended object and use them to derive the orbital (or translational) equations of motion for systems of gravitating bodies.  We consider in particular the case of a binary system containing a point particle (body 1) and an extended deformable star (body 2), working to quadrupolar order in the star's multipole series.  We also discuss an action principle formulation of the orbital dynamics, the process of energy transfer between the gravitational field and the deformable body, and the evolution of the body's spin due to tidal torques.  Finally, we discuss the evolution of the body's tidally induced quadrupole moment in the adiabatic limit.

\subsection{Field equations}\label{newtfield}

In Newtonian physics, the scalar potential $\phi(t,\bm{x})$ obeys the Poisson equation,
\be\label{poisson}
\nabla^2\phi=4\pi\rho,
\ee
with $\rho(t,\bm{x})$ being the rest mass density of matter.  The influence of the gravitational field on matter is described by the test particle acceleration $\ddot{x}^i=-\doe_i\phi$, or more generally, by Euler's equation supplemented by the continuity equation (the conservation of mass),
\be \label{euler}
\doe_t(\rho v^i)+\doe_j(\rho v^i v^j + t^{ij})=-\rho\doe_i\phi,
\ee
\be \label{continuity}
\dot{\rho}+\doe_i (\rho v^i)=0,
\ee
with $v^i(t,\bm{x})$ being the matter's velocity field, and $t^{ij}(t,\bm{x})$ the material stress tensor.  Together, Eqs.~(\ref{poisson}-\ref{continuity}) provide a complete description of Newtonian gravitational interactions.  However, they do not in general form a closed set of evolution equations for the fields $\phi$, $\rho$, $v^i$, and $t^{ij}$; one needs also to specify the matter's internal dynamics, in particular concerning the stress tensor $t_{ij}$.  (In the simplest cases, one can fix $t^{ij}$ by an algebraic equation, like $t^{ij}=0$ for `dust' or $t^{ij}=p\delta^{ij}$ with $p(\rho)$ being the pressure for an isentropic perfect fluid.)  Still, one can derive many useful results, like the form of the translational equations of motion for a system of gravitating bodies, while leaving the matter's internal dynamics unspecified.

\subsection{Multipole moments}\label{newtmult}

We consider $N$ isolated celestial bodies, i.e. regions of space containing matter ($\rho\ne0$) surrounded by regions of vacuum ($\rho=0$), and label the bodies by an index $A$, with $1\le A\le N$.  The potential that is locally generated by body $A$, which we will call the internal potential $\phi_A^{\rm{int}}$, is given by the standard solution to (\ref{poisson}) as an integral over the volume of the body:
\be\label{phiintint}
\phi_A^{\rm{int}}(t,\bm{x})=-\int_A d^3x'\,\frac{\rho(t,\bm{x}')}{|\bm{x}-\bm{x}'|}.
\ee
We can express this potential as a multipole series around a (moving) point $x^i=z_A^i(t)$ by using the Taylor series
\be\label{ntaylor}
\frac{1}{|\bm{x}-\bm{x}'|}=\sum_{l=0}^\infty\frac{(-1)^l}{l!}(x'-z_A)^L\doe_L\frac{1}{|\bm x - \bm z_A|},
\ee
where $L=a_1 a_2 \ldots a_l$ is a spatial multi-index, denoting here tensorial powers of the vector $(x'-z_A)^i$ and of the operator $\doe_i$ (cf. Eq.~(\ref{tensorpower})).  Using the Taylor series (\ref{ntaylor}) in (\ref{phiintint}) allows us to write the internal potential, for points $x^i$ exterior to the body, in the form
\be\label{phiintexp}
\phi_A^{\rm{int}}(t,\bm x)=-\sum_{l=0}^\infty\frac{(-1)^l}{l!}M_A^L(t)\doe_L\frac{1}{|\bm{x}-\bm{z}_A(t)|},
\ee
with
\be\label{defnMAL}
M_A^L(t)=\int_A d^3x\;\rho(t,\bm{x})\left[x-z_A(t)\right]^{<L>}.
\ee
The quantities $M_A^L$ are the mass multipole moments of the body about the worldline $z_A^i(t)$.  They are symmetric, trace-free spatial tensors of varying order $l$, $M_A^L=M_A^{a_1 a_2 \ldots a_l}$.  The STF property follows from the identity (\ref{STFcontract}) and the fact that $\doe_L|\bm x|^{-1}$ is an STF tensor, because partial derivatives commute, and because $\doe_{jj}|\bm x|^{-1}=\nabla^2|\bm x|^{-1}=0$.  As the multipole moments $M_A^L$ are contracted with the STF tensors $\doe_L|\bm{x}-\bm{z}^A|^{-1}$ in (\ref{phiintexp}), only their STF parts will contribute to the potential $\phi_A^{\rm{int}}$; this is why the STF projection (denoted by angular brackets) has been included in the definition of the multipole moments in (\ref{defnMAL}).

The leading-order terms in the multipole series (\ref{phiintexp}) can be written more explicitly as
\be\label{phiintfew}
\phi_A^{\rm{int}}=-\frac{M_A}{|\bar{\bm x}|}-\frac{1}{2} Q_A^{ij}\doe_{ij}\frac{1}{|\bar{\bm x}|}+\ldots
\ee
with
\bse\label{defnmult}
\bea
M_A&=&\int_A d^3x\;\rho, \label{defnM}
\\
0=M_A^i&=&\int_A d^3x\;\rho\;\bar{x}^i, \label{defnCoM}
\\
Q_A^{ij}=M_A^{ij}&=&\int_A d^3x\;\rho\;\left(\bar x^i\bar x^j-\frac{1}{3}|\bar{\bm x}|^2\delta^{ij}\right),\phantom{yoyoyo}\label{defnQ}
\eea
\ese
and $\bar x^i=x^i-z_A^i$.  First is the monopole term ($l=0$), generated by the total mass of the body $M_A$ (\ref{defnM}), and giving rise to a Coulomb-type potential in (\ref{phiintfew}).  We have omitted the dipole term ($l=1$) in (\ref{phiintfew}) because $M_A^i$ can always be made to vanish by choosing the point $z_A^i(t)$ about which the multipole expansion is centered to be fixed to the body's center of mass position:
\be\label{ncom}
z_A^i(t)=\frac{1}{M_A}\int d^3x\;\rho(t,\bm x)\;x^i,
\ee
which is equivalent to (\ref{defnCoM}).  Next comes the $l=2$ term involving the body's mass quadrupole tensor $Q_A^{ij}(t)$ (\ref{defnQ}), which we have renamed $M_A^{ij}\to Q_A^{ij}$ to accord with common convention.  The higher-order terms in the multipole series are suppressed by increasing powers of the finite size parameter $R/|\bar{\bm{x}}|$, where $R$ is the size of the body and $|\bm x|$ is a typical separation.

We note that the continuity equation (\ref{continuity}) implies the constancy of the body's total mass, $\dot{M}_A=0$.  Similarly, the Euler equation (\ref{euler}) and the vanishing of the mass dipole (\ref{defnCoM}) determine the translational equation of motion for the body's worldline $\bm z_A(t)$, as discussed in Sec.~\ref{neoms} below.  Equations (\ref{euler}) and (\ref{continuity}) do not, however, fully determine the evolution of the quadrupole and higher multipoles, which will depend on the body's internal dynamics.

\subsection{Tidal moments}\label{newttidal}

Having described the potential generated by an isolated body with its multipole moments, we can similarly describe the potential felt by the body with tidal moments.  Given a collection of several bodies indexed by $B$, each giving rise its own intrinsic potential of the form (\ref{phiintexp}), we define the external potential felt by a given body $A$ to be the sum of the potentials due to all the other bodies:
\be\label{defphiext}
\phi_A^{\rm{ext}}=\sum_{B\ne A}\phi_B^{\rm{int}}
\ee
The body's tidal moments\footnote{
The subscript $\rm{g}$, standing for `global,' has been included here to avoid confusion with tidal moments introduced in our post-Newtonian treatment below. We introduce there a set of body-frame tidal moments $G_A^L$, defined in an accelerated reference frame attached to the body, and a set of global-frame tidal moments $G_\gA^L$, defined in an (asymptotically) inertial frame.  The tidal moments defined in (\ref{defnG}) coincide with the latter at Newtonian order.  While we have chosen to work exclusively in an inertial frame in our Newtonian treatment here, an analogous Newtonian treatment that uses accelerated frames can be found in Sec.~III of Ref.~\cite{DSX2}.}
$G_\gA^L(t)$ are then defined as coefficients in the Taylor expansion of the external potential about the center-of-mass position $z_A^i$:
\be\label{phiext}
\phi_A^{\rm{ext}}(t,\bm{x})=-\sum_{l=0}^\infty \frac{1}{l!}\, G_\gA^L(t) \left[x-z_A(t)\right]^L,
\ee
\be
G_\gA^L(t)=-\left.\doe_L\phi_A^{\rm{ext}}(t,\bm{x})\right|_{\bm{x}=\bm{z}_A(t)}. \label{defnG}
\ee
Like the multipole moments, the tidal moments are STF tensors, $G_\gA^L=G_\gA^{<L>}$, as can be seen from the definition (\ref{defnG}) and the fact that $\nabla^2\phi_A^{\rm{ext}}=0$ everywhere outside the bodies $B\ne A$.  We see that $G_\gA$ is simply (minus) the potential at the body's center, and $G_\gA^i$ is the would-be test particle acceleration $-\doe_i\phi_A^{\rm{ext}}$.  For $l\ge2$, the $G_\gA^L$ are higher-order derivatives of the potential that will give rise to tidal forces on a non-spherical body. 

The tidal moments of a body $A$ can be expressed in terms of the multipole moments of the other bodies $B\ne A$ by combining (\ref{phiintexp}), (\ref{defphiext}), and (\ref{defnG}), with the result
\bea\label{nGgArel}
G_\gA^L&=&-\sum_{B\ne A} \left.\doe_L\phi_B^{\rm{int}}(t,\bm{x})\right|_{\bm{x}=\bm{z}_A(t)}
\nnm\\
&=&\sum_{B\ne A}\sum_{k=0}^\infty\frac{(-1)^k}{k!}M_B^K\,\doe_{KL}^{(A)}\frac{1}{|\bm{z}_A-\bm{z}_B|},\phantom{yoyoyo}
\eea
where $\doe_i^{(A)}=\doe/\doe z_A^i$, and $\doe_{KL}=\doe_{b_1}\ldots\doe_{b_k}\doe_{a_1}\ldots\doe_{a_l}$.

\subsection{Translational equations of motion}\label{neoms}

A primary advantage of the language of multipole and tidal moments is that it allows one to take the PDEs (\ref{poisson}-\ref{continuity}) governing the evolution of the fields $\rho$, $v^i$, $t^{ij}$, and $\phi$ and extract from them ODEs for the center-of-mass worldlines $z_A^i(t)$ of a collection of gravitating bodies.  To this end, we consider a body $A$ with multipole moments $M_A^L$ defined by (\ref{defnMAL}), in the presence of an external potential $\phi_A^{\rm{ext}}$ generated by other bodies $B\ne A$ according to (\ref{phiext}) and (\ref{nGgArel}).   The body's translational EoM can be found by applying two time derivatives to the definition of $z_A^i$ in (\ref{ncom}), using the Euler and continuity equations (\ref{continuity}) and (\ref{euler}), and integrating by parts.  The result is an expression for the body's center-of-mass acceleration,
\be
M_A \ddot{z}_A^i=-\int d^3x\;\rho\;\doe_i\phi_A^{\rm{ext}},
\ee
which can be rewritten in terms of the body's multipole and tidal moments by using (\ref{phiext}) and (\ref{defnMAL}):
\bea\label{genneom}
M_A \ddot{z}_A^i&=&\sum_{l=0}^\infty\frac{1}{l!}M_A^LG_\gA^{iL}
\\ \nnm
&=&M_A G_\gA^i+\frac{1}{2}Q_A^{jk}G_\gA^{ijk}+\ldots
\eea
The first term in the second line represents the force that would act on a freely falling test mass $M_A$, while the second term gives the leading-order tidal force.

To render the EoM (\ref{genneom}) fully explicit, one can use the expressions for the tidal moments (\ref{nGgArel}) in terms of the multipole moments and worldlines of the other bodies.  Considering the case of a two-body system $A=1,2$, with body $1$ having only a monopole moment $M_1$, and body $2$ having a monopole $M_2$ and a quadrupole $Q^{ij}\equiv Q_2^{ij}$, we find the following EoMs:
\bea
M_1 \ddot z_1^i&=&M_1M_2\doe^{(1)}_i\frac{1}{|\bm{z}_1-\bm{z}_2|}
+\frac{1}{2}M_1 Q^{jk}\doe^{(1)}_{ijk}\frac{1}{|\bm{z}_1-\bm{z}_2|}
\nnm\\ \nnm
M_2 \ddot z_2^i&=&M_2M_1\doe^{(2)}_i\frac{1}{|\bm{z}_2-\bm{z}_1|}
+\frac{1}{2}Q^{jk}M_1\doe^{(2)}_{ijk}\frac{1}{|\bm{z}_2-\bm{z}_1|}
\eea
Defining the radius and unit vector associated with the relative position,
\be\label{reln}
z^i=z_2^i-z_1^i,\quad r=|\bm{z}|,\quad n^i=z^i/r,
\ee
and using the general identity,
\be\label{doeLri}
\doe_L\frac{1}{r}=(-1)^l(2l-1)!!\frac{n^{<L>}}{r^{l+1}},
\ee
these EoMs can be written as
\bea\label{nEoMs}
M_1 \ddot z_1^i&=&-M_2 \ddot z_2^i
\nnm\\
&=&\frac{M_1M_2}{r^2}n_i+\frac{15 M_1}{2 r^4}Q^{jk}n^{<ijk>}\phantom{yoyo}
\eea
In this form, it is evident that the total momentum $p^i=M_1\dot z_1^i+M_2\dot z_2^i$ is conserved, and that these two EoMs for $z_1^i$ and $z_2^i$ can be traded for the EoM of the relative position $z^i=z_2^i-z_1^i$:
\bea\label{nrelEoM}
\ddot z^i&=&-\frac{M}{r^2}n^i-\frac{15 M}{2 M_2 r^4}Q^{jk}n^{<ijk>}
\\ \nnm
&=&-\frac{M}{r^2}n^i-\frac{3 M}{2 M_2 r^4}Q^{jk}(5n^{ijk}-2\delta^{ij}n^k)
\eea
where $M=M_1+M_2$ is the total mass.  In the second line, we have used Eq.~(\ref{STF3}) to expand $n^{<ijk>}$ and the fact that $Q^{jk}$ is STF.

With Eq.~(\ref{nrelEoM}), we have reduced the description of the binary system's translational dynamics to a single ODE.  Still, we can only solve this ODE if we know the time evolution of the quadrupole moment $Q^{ij}(t)$, which requires a detailed model of body $2$'s interior.  We will describe a simple adiabatic model for $Q^{ij}$ in Sec.~\ref{nadiabatic} below.

\subsection{Action principle}\label{naction}

As for any Newtonian system, a Lagrangian for a collection of gravitating bodies can be constructed from $\mc L=T-U$, with $T$ being the kinetic energy and $U$ the potential energy.  The total kinetic energy receives separate contributions from each body, $T=\sum_AT_A$, and each $T_A$ can be split into a contribution from the center-of-mass motion of the body and an internal contribution:
\bea
T_A&=&\frac{1}{2}\int_A d^3x\; \rho\; v^2=\frac{1}{2}M_A\dot{z}_A^2+T_A^{\rm{int}},
\\
T_A^{\rm{int}}&=&\frac{1}{2}\int_A d^3x\; \rho\; (v-\dot{z}_A)^2.
\eea
Here, we have used the fact that $\int_A d^3x\,\rho\,v^i=M_A\dot z_A^i$, implied by the continuity equation (\ref{continuity}) and the definition of the center-of-mass position $z_A^i$ in (\ref{ncom}).  The gravitational potential energy $U=\sum_AU_A$ can be similarly split into external and internal parts:
\bea
U_A&=&\frac{1}{2}\int_A d^3x\; \rho\; \phi=U_A^{\rm{ext}}+U_A^{\rm{int}},
\\
U_A^{\rm{int}}&=&\frac{1}{2}\int_A d^3x\; \rho\;\phi_A^{\rm{int}},
\\
U_A^{\rm{ext}}&=&\frac{1}{2}\int_A d^3x\; \rho\;\phi_A^{\rm{ext}}
=-\frac{1}{2}\sum_{l=0}^\infty\frac{1}{l!}M_A^L G_\gA^L.\phantom{xxxx}
\eea
In the last line, we have used (\ref{phiext}) to express $\phi_A^{\rm{ext}}$ in terms of the tidal moments and (\ref{defnMAL}) for the definition of the multipole moments.

While the system's total potential energy will also receive non-gravitational contributions from the internal structure of each body, we can lump these contributions, along with $T_A^{\rm{int}}$ and $U_A^{\rm{int}}$ as defined above, into an internal Lagrangian $\mc L_A^{\rm{int}}$ for each body.  We can then write the total Lagrangian for an $N$-body system as
\be\label{gennL}
\mc L=\sum_{A}\left(\frac{1}{2}M_A\dot z_A^2+\frac{1}{2}\sum_{l=0}^\infty\frac{1}{l!}M_A^L G_\gA^L+\mc L_A^{\rm{int}}\right).
\ee
The bodies' center-of-mass worldlines $z_A^i(t)$ enter this Lagrangian through the translational kinetic energy terms and through the external gravitational potential energy terms (via the tidal moments); the internal Lagrangians $\mc L_A^{\rm{int}}$, however, are independent of the worldlines $z_A^i$, by construction.  The $\mc L_A^{\rm{int}}$ will be functions of some set of internal configuration variables $q_A^\alpha$ (and their time derivatives) for each body, which will include e.g.~Euler angles for the orientation of the body, vibrational mode amplitudes, etc., depending on the model of the body's internal structure.  (In full generality, the proper internal configuration variables $q_A^\alpha$ are the fields $\rho$, $v^i$ and $t^{ij}$, subject to (\ref{ncom}) as a constraint.)  The bodies' multipole moments $M_A^L$, for $l\ge 2$, appearing in the gravitational potential energy terms in (\ref{gennL}), will be functions of these same internal variables $q_A^\alpha$.  Together, the $z_A^i$ and $q_A^\alpha$, for all bodies $A$, form a complete set of dynamical variables for the $N$-body system.  Varying the action $S=\int\mc L\,dt$ with respect to the worldlines $z_A^i$ reproduces the translational EoMs (\ref{genneom}).  Determining the evolution of the variables $q_A^\alpha$, and hence the moments $M_A^L$ for $l\ge2$, will require a model for $\mc L_A^{\rm{int}}(q_A^\alpha,\dot q_A^\alpha)$ and $M_A^L(q_A^\alpha)$.

Specializing to the two-body $M_1$-$M_2$-$Q_2$ case and using (\ref{nGgArel}), (\ref{reln}), and (\ref{doeLri}), the Lagrangian (\ref{gennL}) becomes
\be\label{nL}
\mc L=\frac{1}{2}M_1\dot z_1^2+\frac{1}{2}M_2\dot z_2^2+\frac{M_1M_2}{r}-U_Q+\mc L_2^{\rm{int}}.
\ee
where $U_Q$ is the potential energy of the quadrupole-tidal interaction:
\be\label{UQ}
U_Q=-\frac{1}{2}Q^{ij}G_{{\rm{g}},2}^{ij},\qquad\quad G_{{\rm{g}},2}^{ij}=\frac{3M_1}{r^3}n^{<ij>}.
\ee
(We have omitted the internal Lagrangian for body $1$ as it is completely decoupled from the rest of the system.)  Varying this action with respect to $z_1^i$ and $z_2^i$ leads to their EoMs found in (\ref{nEoMs}).  Alternately, varying with respect to their separation $z^i=z_2^i-z_1^i$ gives the relative EoM (\ref{nrelEoM}), while the system's center of mass $(M_1z_1^i+M_2z_2^i)/M$ is found to be cyclic.  Specializing the Lagrangian (\ref{nL}) to the center-of-mass frame, where $M_1z_1^i+M_2z_2^i=0$, gives
\be\label{nLcom}
\mc L=\frac{\mu\dot z^2}{2}+\frac{\mu M}{r}-U_Q+\mc L_2^{\rm{int}},
\ee
with $\mu=M_1M_2/M$ being the reduced mass.

  While leaving the functions $\mc L_2^{\rm{int}}(q_2^\alpha,\dot q_2^\alpha)$ and $Q^{ij}(q_2^\alpha)$ unspecified, we can still use the Lagrangian (\ref{nL}) to write down EoMs for body $2$'s internal configuration variables $q_2^\alpha$:
\be\label{qEoM}
\frac{d}{dt}\frac{\doe\mc L_2^{\rm{int}}}{\doe\dot q_2^\alpha}=\frac{\doe \mc L_2^{\rm{int}}}{\doe q_2^\alpha}
+\frac{1}{2}G_{{\rm{g}},2}^{ij}\frac{\doe Q^{ij}}{\doe q_2^\alpha}.
\ee
which will be useful in the next subsection.

\subsection{Energy}\label{nEnergy}

Continuing to specialize to the two-body $M_1$-$M_2$-$Q_2$ case, we can construct a conserved energy for the system from
\be
E=T+U=\frac{\doe\mc L}{\doe\dot z_1^i}\dot z_1^i+\frac{\doe\mc L}{\doe\dot z_2^i}\dot z_2^i
+\frac{\doe\mc L}{\doe\dot q_2^\alpha}\dot q_2^\alpha-\mc L,
\ee
with summation over $\alpha$ implied.  Using the (CoM-frame) Lagrangian (\ref{nLcom}), we find
\be\label{nE}
E=\frac{\mu \dot z^2}{2}-\frac{\mu M}{r}+U_Q+E_2^{\rm{int}},
\ee
where $U_Q$ is given by (\ref{UQ}), and the internal energy of body $2$ is given by
\be\label{E2intq}
E_2^{\rm{int}}(q_2^\alpha,\dot q_2^\alpha)=\frac{\doe\mc L_2^{\rm{int}}}{\doe\dot q_2^\alpha}\dot q_2^\alpha-\mc L_2^{\rm{int}}
\ee
This internal energy will generally have several contributions: internal gravitational potential energy, rotational kinetic energy, vibrational kinetic and potential energy, thermal energy, etc.  Nonetheless, the rate at which energy is exchanged between the interior of body $2$ and its surroundings, via the gravitational tidal interaction, is a function only of the orbital separation, $M_1$, and $Q^{ij}$.  Differentiating (\ref{E2intq}) with respect to time and using the EoM (\ref{qEoM}) for the internal variables $q_2^\alpha$, we find
\bea
\dot E_2^{\rm{int}}&=&\dot q_2^\alpha \left(\frac{d}{dt}\frac{\doe\mc L_2^{\rm{int}}}{\doe\dot q_2^\alpha}
-\frac{\doe\mc L_2^{\rm{int}}}{\doe q_2^\alpha}\right)
=\frac{1}{2}G_{{\rm{g}},2}^{ij}\frac{\doe Q^{ij}}{\doe q_2^\alpha}\dot q_2^\alpha
\nnm\\
&=&\frac{1}{2}G_{{\rm{g}},2}^{ij}\dot Q^{ij}=\frac{3M_1}{2r^3}n^{ij}\dot Q^{ij}\label{E2intdot}
\eea
The energy transfer described by (\ref{E2intdot}) is often referred to as tidal heating (see e.g.~\cite{Purdue}).  This expression for the power delivered to the body is valid (in the quadrupolar approximation) regardless of the body's internal dynamics.  Using (\ref{E2intdot}) and the orbital EoM (\ref{nrelEoM}), one can confirm that the binary system's total energy (\ref{nE}) is conserved.

\subsection{Adiabatic approximation}\label{nadiabatic}

While we have thus far left unspecified the internal dynamics for the deformable body 2, which determines the evolution of the quadrupole, we now specialize our analysis to the case where $Q^{ij}(t)$ is adiabatically induced by the tidal field.  This will lead to a closed system of evolution equations for the binary.  In the adiabatic limit, when the body's internal dynamical time scales are much less than the orbital period, the quadrupole will respond to the instantaneous tidal field according to
\be
Q^{ij}(t)=\lambda G_{{\rm{g}},2}^{ij}(t),\label{nQlambda}
\ee
where $\lambda$ is the tidal deformability, and $G_{{\rm{g}},2}^{ab}(t)$ is the tidal tensor given in Eq.~(\ref{UQ}).  As discussed in Sec.~\ref{intronewton} and in more detail in Refs.~\cite{FH,Hetal}, the relation (\ref{nQlambda}) should be valid to $\sim1\%$ for neutron star binaries at GW frequencies $\lesssim$400 Hz. 

The adiabatic evolution of the quadrupole can be incorporated into our action principle (\ref{nLcom}) by taking $Q^{ij}(t)$ to be our lone internal configuration variable ($q_2^\alpha$), and by taking the internal Lagrangian $\mc L_2^{\rm{int}}$ to contain a simple quadratic potential energy cost for the quadrupole:
\bea\label{nLfinal}
\mc L[z^i,Q^{ij}]&=&\frac{\mu\dot z^2}{2}+\frac{\mu M}{r}-U_Q+\mc L_2^{\rm{int}}
\\ \nnm
&=&\frac{\mu\dot z^2}{2}+\frac{\mu M}{r}+\frac{1}{2}G_{{\rm{g}},2}^{ab}Q^{ab}-\frac{1}{4\lambda}Q^{ab}Q^{ab}.
\eea
Varying the action with respect to the orbital separation $z^i$ still gives the orbital EoM (\ref{nrelEoM}), and varying with respect to the quadrupole $Q^{ij}$ reproduces the adiabatic evolution equation (\ref{nQlambda}).  Using Eq.~(\ref{nQlambda}) to replace $Q^{ij}$ and Eq.~(\ref{UQ}) for $G_{{\rm{g}},2}^{ab}$, the Lagrangian can can written solely in terms of $z^i$ as
\be\label{nLr}
\mc L[z^i]=\frac{\mu\dot z^2}{2}+\frac{\mu M}{r}\left(1+\frac{3\lambda M_1}{2r^5 M_2}\right).
\ee
This Lagrangian leads to the orbital EoM (\ref{nrelEoM}) with $Q^{ij}$ replaced by its adiabatic value (\ref{nQlambda}),
\be\label{nAr}
a^i=-\frac{Mn^i}{r^2}\left(1+\frac{9\lambda M_1}{r^5 M_2}\right),
\ee
which shows that the tidal coupling results in an attractive force.  For circular orbits, with $a^i=-r\omega^2n^i$, we find the radius-frequency relationship
\be\label{nromega}
r(\omega)=\frac{M^{1/3}}{\omega^{2/3}}\left(1-\frac{3\lambda M_1\omega^{10/3}}{M_2M^{5/3}}\right),
\ee
to linear order in the tidal deformability $\lambda$.

The internal energy $E^{\rm{int}}_2$ (\ref{E2intq}), in the adiabatic approximation, is given by
\be\label{nadE}
E^{\rm{int}}_2=\frac{1}{4\lambda}Q^{ab}Q^{ab},
\ee
up to a constant, and satisfies the tidal heating equation (\ref{E2intdot}) by virtue of Eq.~(\ref{nQlambda}).  (We should note that there is actually no `heating' taking place here, as this model neglects dissipative effects and is completely conservative.)  Using Eq.~(\ref{nQlambda}), the binary system's total energy $E$ (\ref{nE}) can be written as
\be\label{nadEtot}
E=\frac{\mu\dot z^2}{2}-\frac{\mu M}{r}\left(1+\frac{3\lambda M_1}{2r^5 M_2}\right),
\ee
in the adiabatic model, and is conserved by the orbital EoM (\ref{nAr}).  Then, using $\dot z^2=r^2\omega^2$ and Eq.~(\ref{nromega}), we find
\be\label{nEomega}
E(\omega)=-\frac{\mu}{2}(M\omega)^{2/3}\left(1-\frac{9\lambda M_1\omega^{10/3}}{M_2M^{5/3}}\right),
\ee
for the circular-orbit energy-frequency relationship.

\subsection{Spin}\label{nspin}

In anticipation of the 1PN treatment of the binary's orbital dynamics, in which a body's angular momentum (or spin) has a direct influence on the orbit, it will be useful to discuss the evolution of an extended body's spin at Newtonian order.  The spin of a body $A$ about its CoM worldline $z_A^i$ is defined by
\be\label{defnS}
S_A^a(t)=\epsilon^{abc}\int d^3x\,\rho(t,\bm{x}) \left[x^b-z_A^b(t)\right]v^c(t,\bm{x}),
\ee
with $\rho$ being the mass density and $v^c$ the velocity field.  Taking a time derivative of this equation, using the Euler and continuity equations (\ref{euler}) and (\ref{continuity}) and the definition of $z_A^i$ in (\ref{ncom}), and integrating by parts, we find
\bea\label{nSdotgen}
\dot S_A^a&=&-\epsilon^{abc}\int d^3x\,\rho\,(x^b-z_A^b)\,\doe_c \phi_A^{\rm{ext}}
\nnm\\
&=&\epsilon^{abc}\sum_{l=0}^\infty \frac{1}{l!} M_A^{bL} G_\gA^{cL}.
\eea
In the second line, we have used the definitions of $M_A^L$ and $G_\gA^L$ in (\ref{defnMAL}) and (\ref{defnG}).  This formula gives the torque on the body due to tidal forces.  As it is not directly relevant to our purposes, we will not discuss a Lagrangian formulation of the Newtonian rotational dynamics.

Applying Eq.~(\ref{nSdotgen}) to the $M_1$-$M_2$-$Q_2$ system, we find that the tidal torque on body 2 is given by
\be\label{nspineom}
\dot{S}_2^a=\epsilon^{abc}Q^{bd}G_{{\rm{g}},2}^{cd}
\ee
with the tidal tensor $G_{{\rm{g}},2}^{cd}$ given by (\ref{UQ}).  Eq.~(\ref{nspineom}) is valid (in the quadrupolar approximation) regardless of the internal dynamics of body 2.  In the special case of an adiabatically induced quadrupole, as in Eq.~(\ref{nQlambda}), we find
\be\label{nadspin}
\dot{S}^a=\lambda\epsilon^{abc}G_{{\rm{g}},2}^{bd}G_{{\rm{g}},2}^{cd}=0,
\ee
so that the spin is conserved.

\section{Post-Newtonian tidal interactions}\label{pnti}

\subsection{Overview}\label{ov}

The Newtonian theory of gravity arises as a limiting case of general relativity (GR).  In the limit of small source velocities and weak gravity, the spacetime metric of GR takes the form
\be\label{nmetric}
ds^2=-\left(1+\frac{2\phi}{c^2}\right)c^2dt^2+\delta^{ij}dx^idx^j+O(c^{-2}),
\ee
with $\phi(t,\bm{x})$ being the Newtonian potential.  This expression represents a perturbation expansion of the theory with $1/c^2$ playing the role of a formal expansion parameter.  At leading order in $1/c^2$, Einstein's equation and covariant stress-energy conservation for the metric (\ref{nmetric}) reproduce the Poisson, Euler, and continuity equations (\ref{poisson}-\ref{continuity})---the basic equations of Newtonian gravity.

The first post-Newtonian (1PN) approximation to GR continues this perturbation expansion to next-to-leading order in $1/c^2$.  The 1PN metric can be written as
\bea
ds^2&=&-\left[1+\frac{2\phi}{c^2}+\frac{2}{c^4}(\phi^2+\psi)\right]c^2dt^2
+\frac{2\zeta^i}{c^2}dt\,dx^i
\nnm\\
&&+\left(1-\frac{2\phi}{c^2}\right)\delta^{ij}dx^i dx^j +O(c^{-4}),
\eea
(cf.~Weinberg \cite{Weinberg}), with two new degrees of freedom appearing: a 1PN (three-)vector potential $\zeta^i(t,\bm{x})$ (often called the gravito-magnetic potential), and a 1PN scalar potential $\psi(t,\bm{x})$.  Following Refs.~\cite{BD,DSX1} (apart from
a change of sign) we shall work with the single scalar potential single scalar potential $\Phi(t,\bm{x})$ which has $\phi$ and $\psi$ as its Newtonian- and 1PN-order parts (and hence a hidden $c$-dependence),
\be
\Phi=\phi+c^{-2}\psi+O(c^{-4}),
\ee
so that the metric can be written as
\bea
ds^2&=&-\left[1+\frac{2\Phi}{c^2}+\frac{2\Phi^2}{c^4}\right]c^2dt^2
+\frac{2\zeta^i}{c^2}dt\,dx^i
\nnm\\
&&+\left(1-\frac{2\Phi}{c^2}\right)\delta^{ij}dx^i dx^j +O(c^{-4}).
\label{metric}
\eea

We choose to work here in conformally Cartesian coordinates (see e.g.~\cite{DSX1}), which is already implicit in the form (\ref{metric}) of the metric, and to adopt the harmonic gauge condition: 
\be\label{harmonic}
\doe_\mu(\sqrt{-g}g^{\mu\nu})=0\quad\Leftrightarrow\quad 4\dot{\Phi}+\doe_i\zeta^i=O(c^{-2}).
\ee
In this gauge, one finds that Einstein's equation for the metric, at next-to-leading order in $1/c^2$, is equivalent to the following linear field equations for the potentials:
\bse\label{poteq}
\bea
\nabla^2\Phi&=&4\pi T^{tt}+c^{-2}\left(4\pi T^{ii}+\ddot{\Phi}\right)+O(c^{-4}) ,\phantom{yoyo} \label{phieq}
\\
\nabla^2\zeta^i&=& 16\pi T^{ti} + O(c^{-2}), \label{zetaeq}
\label{psieq}
\eea
\ese
where $T^{\mu\nu}$ are the contravariant components of the stress-energy tensor in the $(t,x^i)$ coordinate system.  Note that the $T^{tt}$ component must include both $O(c^0)$ Newtonian and $O(c^{-2})$ post-Newtonian contributions, while the components $T^{ti}$ and $T^{ij}$ and the quantity $\ddot\Phi$ are needed only to Newtonian order.  The influence of the gravitational field on matter is governed by covariant stress-energy conservation: $\nabla_\mu T^{\mu\nu}=0$; the 1PN-expanded form of this equation can be found in Appendix D of RF \cite{RF}.

In the remainder of this section, we review the treatment of tidal interactions within this first-post-Newtonian framework.  We employ a formalism, originally developed by DSX \cite{DSX1,DSX2} and later expounded upon by RF \cite{RF}, that uses multiple coordinate systems to describe the global motion and local structure of extended bodies.  We attempt to present here the primary ingredients and broad logical flow of this formalism, which are essential for properly interpreting the results stated in Sec.~\ref{results} above.

We begin in Sec.~\ref{body} by presenting a general solution to the 1PN-order Einstein equations (\ref{poteq}) which gives the spacetime metric in a vacuum region surrounding an astronomical body $A$.  The solution (\ref{potA}) is parametrized by and defines the body's multipole moments and tidal moments.  The mass and current multipole moments $M_A^L$ and $S_A^L$ characterize the body's internal structure, and the gravito-electric and -magnetic tidal moments $G_A^L$ and $H_A^L$ characterize the external gravitational fields felt by the body.

In Sec.~\ref{ct} we discuss the gauge freedom in the 1PN metric, summarized by the parametrization of a general 1PN coordinate transformation in Eq.~(\ref{xt}).  In 1PN celestial mechanics, it is advantageous to use and transform between two types of coordinate systems: global coordinates $(t,x^i)$ used to describe the motion of multiple bodies, and body-adapted coordinates $(s_A,y_A^i)$ used in the local description of a given body $A$.  We discuss how to fix all 1PN coordinate freedom in the body-adapted coordinates $(s_A,y_A^i)$ by enforcing the body-frame gauge conditions (\ref{bfgauge}).  The body-frame multipole moments $M_A^L(s^A)$ and $S_A^L(s^A)$---the moments defined by the multipole expansions in Sec.~\ref{body} using the body-adapted coordinates---then become unique and meaningful descriptors of a body's internal structure.

In Sec.~\ref{global}, we discuss the form of the metric in the global coordinates $(t,x^i)$.  It is written in terms of a set of global-frame multipole moments $M_\gA^L(t)$ and $Z_\gA^{iL}(t)$ and tidal moments $G_\gA^L(t)$ and $Y_\gA^{iL}(t)$ for each body $A$, which differ from the body-frame moments.  The relationship between the global- and body-frame moments is determined by the transformation (\ref{xt}) between the global and body-adapted coordinates; the moment transformation formulae are presented in full detail in Appendix \ref{atm}.  The functions parameterizing the coordinate transformation, or the worldline data $\mc{D}_A$ (\ref{DA}), are seen to take on the role of configuration variables for body $A$ with respect to the global frame.  Among other things, they determine the body's center-of-mass worldline, $x^i=z_A^i(t)$.

In Sec.~\ref{loms}, we discuss the 1PN single-body laws of motion (\ref{lom_}) which govern the evolution of a body's mass monopole $M_A$, mass dipole $M_A^i$, and current dipole (or spin) $S_A^i$.  These laws reflect the conservation of energy, momentum, and angular momentum and can be derived from stress-energy conservation at 1PN order \cite{DSX2}, or equivalently, from Einstein's equation at 2PN order \cite{RF}.  A body's translational equation of motion---an ODE for its global-frame center-of-mass worldline $z_A^i(t)$---can be deduced from the law of motion for its body-frame mass dipole.  The result is an expression for the acceleration $\ddot z_A^i$, for each member $A$ of an $N$-body system, written in terms of the body-frame multipole moments $M_A^L$ and $S_A^L$ and global-frame worldlines $z_A^i$ of all the bodies $A$.

Finally, in Sec.~\ref{truncate}, we specialize our discussion to the case of a two-body system with a body 1 having only a mass monopole $M_1$, and a body 2 having a mass monopole $M_2$, a mass quadrupole $Q_2^{ij}\equiv Q^{ij}$, and a spin $S^i_2\equiv S^i$.  We present and discuss the explicit forms of the evolution equations for the moments $M_1$, $M_2$, and $S^i$ and the worldlines $z_1^i$ and $z_2^i$, which depend only on these quantities and $Q^{ij}$.

\subsection{1PN multipole and tidal moments}\label{body}

In Sec.~\ref{newton}, we defined the Newtonian mass multipole moments $\n M_{A}^{L}$ (called simply $M_A^L$ there) as integrals over the body's mass distribution (\ref{defnMAL}).  In so doing, we implicitly assumed that the Newtonian Poisson equation (\ref{poisson}) was valid in all space, including the interior of the body.  A similar approach can be taken at 1PN order, defining 1PN-accurate multipole moments as integrals over a body's stress-energy distribution (as in (\ref{stressint}) below), assuming that the 1PN field equations (\ref{poteq}) are valid in all space.  This was the approach taken in the original DSX formalism.  

As stressed by RF, one can also define a body's 1PN multipole moments without requiring the validity of the 1PN field equations in the interior of the body.  Instead, one need only impose the field equations in a vacuum buffer region $\mc{B}_A$, a region of finite extent enclosed between two coordinate spheres centered on the body; the moments can then be defined through the multipole expansion of the 1PN metric in the region $\mc{B}_A$.  This allows one to consider objects with strong internal gravity, like neutron stars and black holes, as long as there exists a region $\mc{B}_A$ exterior to the object where gravity is sufficiently weak and quasi-static for the 1PN field equations to be valid.

Taking the latter approach, we assume the existence of a local coordinate system $(s_A,y_A^i)$, in the vicinity of the body $A$, having the following properties:  (i) The range of the coordinates includes the product of the open ball $|\bm{y}_A|<r_2$, for some finite radius $r_2$, with an open interval of time $(s_A^1,s_A^2)$.  (ii) There exists a spatial region $\mc{W}_A$ (the worldtube) of the form $|\bm{y}_A|<r_1$ that contains all the body's stress-energy and/or regions of strong gravity. (iii) In the buffer region $\mc{B}_A$ $(r_1<|\bm{y}_A|<r_2)$,
the coordinates $(s_A,y_A^i)$ are conformally Cartesian and harmonic, and the metric takes the 1PN form (\ref{metric}), with potentials $\Phi_A(s_A,\bm y_A)$ and $\zeta_A^i(s_A,\bm y_A)$ satisfying the 1PN vacuum field equations:
\bea
\nabla^2\Phi_A&=&c^{-2}\ddot\Phi+O(c^{-4}), \label{Avaceq}
\nnm\\
\nabla^2\zeta_A^i&=&O(c^{-2}).
\eea
Under these assumptions, RF showed that the general solution for the potentials in $\mc{B}_A$ is of the form
\begin{widetext}
\bse\label{potA}
\bea
\Phi_A(s_A,\bm y_A)&=&-\sum_{l=0}^\infty\frac{1}{l!}\bigg\{
(-1)^l M_A^L(s_A)\doe_L\frac{1}{|\bm{y}_A|}
+G_A^L(s_A)y_A^L                                      \label{phiA}
\\ \nnm
&&\phantom{ -\sum_{l=0}^\infty\frac{1}{l!}\bigg\{ }
+\frac{1}{c^2}\bigg[
\frac{(-1)^l(2l+1)}{(l+1)(2l+3)}\dot{\mu}_A^L(s_A)\doe_L\frac{1}{|\bm{y}_A|}
+\frac{(-1)^l}{2}\ddot{M}_A^L(s_A)\doe_L|\bm{y}_A|
\nnm\\
&&\phantom{ -\sum_{l=0}^\infty\frac{1}{l!}\bigg\{+\frac{1}{c^2}\bigg[ }
-\dot{\nu}_A^L(s_A)y_A^L+\frac{1}{2(2l+3)}\ddot{G}_A^L(s_A)y_A^{jjL}\bigg]\bigg\}+O(c^{-4}),
\\
\zeta_A^i(s_A,\bm y_A)&=&-\sum_{l=0}^\infty\frac{1}{l!}\bigg\{(-1)^l\,Z_A^{iL}(s_A)\,\doe_L\frac{1}{|\bm{y}_A|}+Y_A^{iL}(s_A)\,y_A^L\bigg\}+O(c^{-4}),\label{zetaA}
\eea
\ese
with
\bse\label{ZYA}
\bea
Z_A^{iL}(s_A)&=&\frac{4}{l+1}\dot{M}_A^{iL}(s_A)-\frac{4l}{l+1}\epsilon^{ji<a_l}S_A^{L-1>j}(s_A)+\frac{2l-1}{2l+1}\delta^{i<a_l}\mu_A^{L-1>}(s_A)+O(c^{-4}),\label{ZA}
\\
Y_A^{iL}(s_A)&=&\nu_A^{iL}(s_A)+\frac{l}{l+1}\epsilon^{ji<a_l}H_A^{L-1>j}(s_A)-\frac{4(2l-1)}{2l+1}\dot{G}_A^{<L-1}\delta^{a_l>i}(s_A)+O(c^{-4}).\label{YA}
\eea
\ese
\end{widetext}

The potentials are parametrized by the following sets of (multi-index) spatial tensors, which are STF on all their indices, and which are functions only of the time coordinate $s_A$.  First, $M_A^L(s_A)$, with $l\ge0$, are the body's mass multipole moments, which are defined with 1PN-accuracy.  Next are the current multipole moments, $S_A^L(s_A)$, with $l\ge1$, needed only to Newtonian accuracy.  Together, the mass and current multipole moments contain all the information about the body's internal structure that is encoded in the gravitational field it produces (at 1PN order).  They are associated with the contributions to the potentials that appear to diverge as $|\bm{y}_A|\to0$, which can be referred to as the internal contributions.  Also associated with such parts of the potentials are the internal gauge moments $\mu_A^L$ $(l\ge0)$, so called because they contain no gauge-invariant information about the body.  

Associated with the parts of the the potentials that appear to diverge as $|\bm{y}_A|\to\infty$ (the external parts) are the tidal moments: the gravito-electric tidal moments, $G_A^L(s_A)$, with $l\ge0$, are defined with 1PN accuracy (like $M_A^L$), and the gravito-magnetic tidal moments $H_A^L(s_A)$, with $l\ge 1$, are defined with Newtonian accuracy (like $S_A^L$).  The tidal moments contain information about gravitational fields generated by external sources and about inertial effects associated with the motion of the local coordinate system.  Finally, there are the tidal gauge moments $\nu_A^L$, defined for $l\ge1$.

The tensors $Z_A^{iL}(s_A)$ and $Y_A^{iL}(s_A)$ appearing in the gravito-magnetic potential (\ref{zetaA}) have been defined as useful shorthands for the expressions in (\ref{ZYA}).  Unlike all the other moments just introduced, they are not STF on all their indices, but they are STF on their last $l$ indices (i.e.~on all but the first index).  Eqs.~(\ref{ZYA}) in fact represent their unique decompositions in terms of fully STF tensors; the `inverse' relations are
\bse\label{Zto_}
\bea
S_A^L&=&-\frac{1}{4}Z_A^{jk<L-1}\epsilon^{a_l>jk},\label{ZtoS}
\\
\mu_A^L&=&Z_A^{jjL},\label{Ztomu}
\\
\dot{M}_A^{iL}&=&-\frac{l+1}{4}Z_A^{<iL>},\label{ZtoMdot}
\eea
\ese
and
\bse\label{Yto_}
\bea
H_A^L&=&Y_A^{jk<L-1}\epsilon^{a_l>jk},\label{YtoH}
\\
\nu_A^L&=&Y_A^{<L>},\label{Ytonu}
\\
\dot{G}_A^{L}&=&-\frac{1}{4}Y_A^{jjL}.\label{YtoGdot}
\eea
\ese
The relations (\ref{ZtoMdot}) and (\ref{YtoGdot}) are implied by the harmonic gauge condition (\ref{harmonic}).

In the case where the 1PN field equations (\ref{poteq}) are in fact valid in the interior of the body, the mass and current multipole moments can be defined by integrals over the stress-energy distribution in the volume of the body, as in DSX \cite{DSX2}:
\bse\label{stressint}
\bea
M_A^L&=&\int_Ad^3y_A \bigg\{y_A^{<L>} T^{tt}  \label{siM}
\\ \nnm
&& + \frac{1}{c^2}\bigg[y_A^{<L>}T^{jj}+\frac{1}{2(2l+3)}y^{jj<L>}_A \ddot T^{tt}
\nnm\\
&& -\frac{4(2l+1)}{(l+1)(2l+3)}y_A^{<jL>}\dot{T}^{tj}\bigg]\bigg\}+O(c^{-4}),
\nnm\\
S_A^L&=&\!\int_A\!d^3y_A\;\epsilon^{jk<a_l} y_A^{L-1>j} T^{tk} +O(c^{-2}).\phantom{yoyoyo}\label{siS}
\eea
\ese
When considering a body with strong internal gravity, its interior cannot be modeled by a 1PN stress-energy distribution, and such integrals cannot be defined.  Instead, we rely on the multipole expansion of the potentials (\ref{potA}) in the buffer region $\mc{B}_A$ to define the multipole moments $M_A^L$ and $S_A^L$, as well as the tidal moments $G_A^L$ and $H_A^L$.  Appendix E of RF \cite{RF} demonstrates the sufficiency of this method of definition by giving explicit formulae for the moments in terms of surface integrals of the potentials in $\mc{B}_A$.

\subsection{Coordinate transformations and body-frame gauge conditions}\label{ct}

The 1PN metric (\ref{metric}) harbors residual coordinate freedom not fixed by the conformally Cartesian and harmonic gauge conditions.  As a result, the multipole and tidal moments defined in the last section (not just the `gauge moments' $\mu_A^L$ and $\nu_A^L$, but rather all of the moments) are not unique and will vary with the choice of coordinates.  To define a unique set of multipole moments for a given body, one must further specialize the body-frame coordinates.  Thus we turn now to a discussion of 1PN coordinate transformations.

In RF \cite{RF}, it was shown that the most general transformation between two harmonic coordinate systems $(s,y^i)$ and $(t,x^i)$ in which the metric takes the 1PN form (\ref{metric}) can be written as
\begin{widetext}
\bse\label{xt}
\bea\label{x}
x^i(s,\bm y)&=&y^i+z^i(s)+\frac{1}{c^2}\bigg\{
\left[\frac{1}{2}\dot{z}^{kk}(s)\delta^{ij}-\dot{\alpha}(s)\delta^{ij}+\epsilon^{ijk}R^k(s)+\frac{1}{2}\dot{z}^{ij}(s)\right]y^j
\nnm\\
&&\phantom{y^i+z^i(s)+\frac{1}{c^2}\bigg\{}
+\left[\frac{1}{2}\ddot{z}^i(s)\delta^{jk}-\ddot{z}^k(s)\delta^{ij}\right]y^{jk}\bigg\}+O(c^{-4}),
\\
\label{t}
t(s,\bm y)&=&s+\frac{1}{c^2}\left[\alpha(s)+\dot{z}^j(s)y^j\right]
+\frac{1}{c^4}\left[\beta(s,\bm y)+\frac{1}{6}\ddot{\alpha}(s)y^{jj}+\frac{1}{10}\dddot{z}_A^j(s)y^{jkk}\right]+O(c^{-6}),
\eea
\ese
\end{widetext}
being parametrized by the following functions.  The vector $z^i(s)$ provides a time-dependent translation between the spatial coordinates and is defined with 1PN accuracy.  Each defined with Newtonian accuracy\footnote{
Though an $O(c^{-2})$ contribution to $\alpha(s)$ would contribute at $O(c^{-4})$ in Eq.~(\ref{t}), this contribution can be absorbed into the function $\beta(s,\bm y)$.}
 are the rotation vector $R_i(s)$, and the functions $\alpha(s)$ and $\beta(s,\bm{y})$ which transform the time coordinate.  All of these may be arbitrary functions of their arguments (within the bounds of their post-Newtonian scaling), except that $\beta(s,\bm y)$ must be harmonic, $\nabla^2\beta=0$, in order to preserve the harmonic gauge condition.

In the treatment of the $N$-body problem, we will make use of one global coordinate system $(t,x^i)$ and one body-adapted coordinate system $(s_A,y_A^i)$ for each body $A$.  The global coordinates (described further in the next section) are used to track the bulk motion of all the bodies, while the body-adapted coordinates are used in the local description of each body---in particular, to define their body-frame multipole and tidal moments.  The transformation between the $(t,x^i)$ and $(s_A,y_A^i)$ coordinates will take the form (\ref{xt}), with different 'worldline data' functions,
\be\label{DA}
\mc{D}_A=\{z_A^i(s^A),R_A^i(s_A),\alpha_A(s_A),\beta_A(s_A,y_A^i)\}
\ee
for each body $A$.  These functions may be viewed as configuration variables for the body-adapted frame, specifying its position, orientation, etc.~relative to the global frame.

In order to uniquely define the body-frame multipole and tidal moments, we must fix some of the remaining gauge freedom in the body-adapted coordinates $(s_A,y_A^i)$.  Here, we will fix all remaining gauge freedom in the body-adapted coordinates (in the bodies' buffer regions), which will also uniquely determine the gauge moments.  It was shown in RF that this can be always be accomplished by imposing the following conditions, which define the body-adapted gauge:
\bse\label{bfgauge}
\bea
&M_A^i(s_A)=0&\label{masscenter}
\\
&R_A^i(s_A)=0&\label{coriolis}
\\
&G_A(s_A)=\mu_A(s_A)=0&\label{zeropotential}
\\
&\mu_A^L(s_A)=\nu_A^L(s_A)=0,\;\;l\ge1&\label{munuzero}
\eea
\ese
Eq.~(\ref{masscenter}), setting the body-frame mass dipole $M_A^i$ to zero, fixes the body's center of mass-energy to the origin of the spatial coordinates $y_A^i=0$ to 1PN order.  Setting the rotation vector $R_A^i$ to zero in Eq.~(\ref{coriolis}) fixes the orientation of the body-frame spatial axes to those of the global frame.\footnote{
In place of the condition (\ref{coriolis}), RF chose to set the gravito-magnetic dipole tidal moment $H_A^i(s_A)$ to zero, which cancels leading-order Coriolis forces in the body-adapted frame and requires a non-zero value of the rotation vector $R_A^i(s_A)$.  While this more completely effaces external gravitational and inertial effects in the body frame, Eq.~(\ref{coriolis}) leads to more simplifications in calculations.  The effects of these differing gauge choices cancel in all final results. \label{RHfoot}}  If the extended body $A$ were replaced by a freely falling observer at $y_A^i=0$ (assuming an extension of the body-frame coordinates to $y_A^i=0$), Eq.~(\ref{zeropotential}) would ensure that the time coordinate $s^A$ measures their proper time.  Finally, the fact that all the gauge moments can always be set to zero by a coordinate transformation, as in (\ref{munuzero}), shows that they are pure gauge degrees of freedom.

We can think of the body-adapted coordinates as defining the body's local asymptotic rest frame \cite{ThorneHartle}, in which the effects of external gravitational fields and inertial effects have been removed as much as possible$^4$.  The body-frame moments---the multipole and tidal moments defined by (\ref{potA}) in the body-adapted coordinates---then take on the values that would be measured by a local comoving observer in the body frame.  The body-frame multipole moments $M_A^L$ and $S_A^L$ are the quantities describing the bodies' internal structure that will appear in the final form of the translational equation of motions for an $N$-body system.

\subsection{The global frame}\label{global}

To treat the orbital dynamics of a collection of several bodies $A=1\ldots N$, we consider $N+1$ separate coordinate systems: one body-adapted coordinate system $(s_A,y_A^i)$ for each body $A$, and one global coordinate system $(t,x^i)$.  We take these coordinate systems to have the following properties: (i) For each body $A$, the body-adapted coordinates $(s_A,y_A^i)$ cover the body's buffer region $\mc{B}_A$ and satisfy all the assumptions and gauge conditions outlined in Secs.~\ref{body} and \ref{ct}.  (ii) The bodies' buffer regions $\mc{B}_A$ are non-overlapping.  (iii) The global coordinates $(t,x^i)$ cover the buffer regions of all the bodies as well as the intervening space; i.e.~they cover the region $\mc{B}_{\rm{g}}=\mc{M}\backslash\bigcup_A\mc{W}_A$, the entire spacetime manifold $\mc{M}$ except for the worldtubes.  (iv) In the region $\mc{B}_{\rm{g}}$, the coordinates $(t,x^i)$ are conformally Cartesian and harmonic, and the metric takes the form (\ref{metric}), with potentials $\Phi_\g(t,\bm{x})$ and $\zeta_\g^i(t,\bm{x})$ satisfying the PN vacuum field equations (Eqs.~(\ref{Avaceq}) with $A\to {\rm{g}}$).

The final assumption allows us to write down the following multipole expansion of the global-frame potentials in $\mc{B}_{\rm{g}}$:
\begin{widetext}
\bse\label{potg}
\bea
\Phi_\g(t,\bm{x})&=&-\sum_{A=1}^N\sum_{l=0}^{\infty}\frac{(-1)^{l}}{l!}\left\{M_\gA^L(t) \doe_L\frac{1}{|\bm{x}-\bm{z}_A(t)|}
+\frac{1}{2c^2}\doe_t^2\Big[M_\gA^L(t) \doe_L|\bm{x}-\bm{z}_A(t)|\Big]\right\}+O(c^{-4}),    \label{phig}
\\
\zeta^i_\g(t,\bm{x})&=&-\sum_{A=1}^N\sum_{l=0}^{\infty}\frac{(-1)^l}{l!}Z_\gA^{iL}(t) \doe_L\frac{1}{|\bm{x}-\bm{z}_A(t)|}+O(c^{-2}),      \label{zetag}
\eea
\ese
\end{widetext}
This expansion is analogous to that for the body-frame potentials (\ref{potA}) but has several important differences.  Firstly, the potentials are written as a sum of contributions from each body $A$; this is justified by the linearity of the field equations (\ref{poteq}).  Each such contribution is parametrized by the body's global-frame multipole moments: the mass multipole moments $M_\gA^L(t)$ are fully STF and 1PN-accurate, and the tensors $Z_\gA^{iL}(t)$ are STF on all but the first index and Newtonian-accurate.  Both sets of tensors are defined for $l\ge0$.  As these global-frame moments will only appear in intermediate stages of our calculations, we will not bother decomposing the tensors $Z_\gA^{iL}$ in terms of fully STF current and gauge moments as in the body-frame case (\ref{ZA}).  We have taken the moments $Z_\gA^{iL}$ to satisfy
\be\label{Zgconds}
Z_{g,A}^{<iL>}=-\frac{4}{l+1}\dot M_\gA^{iL},\quad\qquad Z_{g,A}^{jjL}=0.
\ee
The first of these is required by the harmonic gauge condition, and the second is equivalent to setting the would-be global-frame gauge moments $\mu_\gA^L$ to zero.\footnote{
The second condition in (\ref{Zgconds}), along with the fact that the global-frame potentials all vanish as $|\bm x|\to\infty$, reduces the residual gauge freedom in the global-frame metric to the group of post-Galilean transformations (the post-Newtonian Poincar\'e group) \cite{RF}, which are the coordinate transformations given by (\ref{xt}) with $\ddot z^i=\dot R^i=\beta=0$ and $\dot\alpha=\dot z^2/2$.}
The global-frame moments $M_\gA^L(t)$ and $Z_\gA^{iL}(t)$ are distinct from (though related to) the corresponding body-frame moments $M_A^L(s_A)$ and $Z_A^{iL}(s_A)$.

A second important difference with the body frame case is that the multipole expansions appearing here are centered not around the spatial origin $x^i=0$ but around the worldlines $x^i=z_A^i(t)$.  One can check that the potentials as written here still satisfy the harmonic-gauge 1PN field equations in $\mc B_\g$ for any choices of these worldlines.  Below, we will identify the $z_A^i$ with the bodies' center-of-mass worldlines, which appear as parameters in the transformations from body-adapted to global coordinates (\ref{xt}).

Finally, one can note that we have included, in each body's contributions to the potentials, only internal pieces (which appear to diverge as $|\bm{x}-\bm{z}^A|\to0$) and not tidal pieces (which would appear to diverge as $|\bm{x}-\bm{z}^A|\to\infty$).  This makes the global-frame metric tend to the Minkowski metric as $|\bm{x}|\to\infty$, thus eliminating any tidal or inertial forces on the $N$-body system as a whole.  Each body will still experience local tidal fields, but they will arise from the contributions to the potentials generated by the other bodies.

We can introduce a set of global-frame tidal moments for each body $A$ by rewriting the global-frame potentials, in the body's buffer region $\mc B_A$, as
\begin{widetext}
\bse\label{potgtidal}
\bea
\Phi_\g(t,\bm{x})&=&-\sum_{l=0}^{\infty}\frac{1}{l!}\bigg\{(-1)^l M_\gA^L(t) \doe_L\frac{1}{|\bm{x}-\bm{z}_A(t)|}+G_\gA^L(t)[x-z_A(t)]^L        \label{phigtidal}
\\ \nnm
&&\phantom{-\sum_{l=0}^{\infty}\frac{1}{l!}\bigg\{}
+\frac{1}{2c^2}\doe_t^2\left[(-1)^l M_\gA^L(t) \doe_L|\bm{x}-\bm{z}_A(t)|+\frac{1}{2l+3}G_\gA^L(t)[x-z_A(t)]^{jjL}\right]\bigg\}+O(c^{-4})
\\
\zeta^i_{\rm{g}}(t,\bm{x})&=&-\sum_{l=0}^{\infty}\frac{1}{l!}\left[(-1)^lZ_\gA^{iL}(t) \doe_L\frac{1}{|\bm{x}-\bm{z}_A(t)|}+ Y_\gA^{iL}[x-z_A(t)]^L\right]+O(c^{-2}),  \label{zetagtidal}
\eea
\ese
\end{widetext}
Here, we have absorbed the contributions to the potentials from the other bodies $B\ne A$ into tidal terms for body $A$.  This defines the global-frame tidal moments $G_\gA^{L}(t)$ and $Y_\gA^{iL}(t)$.  They can be expressed in terms of the global-frame multipole moments of the other bodies $B\ne A$ and the worldlines $z_A^i$ of all the bodies $A$ by equating the expressions for the potentials in (\ref{potgtidal}) with those in (\ref{potg}); these relations are given in Appendix~\ref{gfmult_gftidal}.

Now, as the global coordinates $x^\mu=(t,x^i)$ and the body-adapted coordinates $y_A^\mu=(s_A,y_A^i)$ are related by the coordinate transformation (\ref{xt}), the metrics in the global and body frames must be related by the tensor transformation law:
\be\label{gtransf}
g^A_{\mu\nu}=\frac{\doe x^\rho}{\doe y_A^\mu}\frac{\doe x^\sigma}{\doe y_A^\nu}g^g_{\rho\sigma}
\ee
This requirement allows one to determine both the parameters of the coordinate transformation between the two coordinate systems (\ref{xt}) and the relationship between the global- and body-frame multipole and tidal moments.  Making use of the form (\ref{metric}) for the metric in terms of the potentials (in both coordinate systems) and the expressions for the body-frame potentials (\ref{potA}) and the global-frame potentials (\ref{potgtidal}), as detailed in RF \cite{RF}, Eq.~(\ref{gtransf}) yields expressions for the body-frame moments in terms of the global-frame moments and the worldline data (or the inverse relations):
\bea
(M_A^L,S_A^L)&\stackrel{\mc{D}_A}{\longleftrightarrow}&(M_\gA^L,Z_\gA^{iL})
\nnm\\ 
(G_A^L,H_A^L)&\stackrel{\mc{D}_A}{\longleftrightarrow}&(G_\gA^L,Y_\gA^{iL})
\eea
These moment transformations are presented in full detail in Appendices~\ref{bfmult_gfmult} and \ref{gftidal_bftidal}.

By combining the transformation formulae for the tidal moments with the body-frame gauge conditions (\ref{bfgauge}), one can solve for and eliminate the worldline data functions $\alpha_A(s_A)$ and $\beta_A(s_A,y_A^j)$.  The only remaining piece of the worldline data $\mc D_A$ (\ref{DA}) is the translation vector $z_A^i(s_A)$.  Recall that the body-frame gauge condition $M_A^i=0$ (\ref{masscenter}), setting the mass dipole to zero, fixes the body's center of mass-energy to the body-frame spatial origin $y_A^i=0$.  Setting $y_A^i=0$ in the coordinate transformation (\ref{xt}) and eliminating $s_A$, we see that $x^i=z_A^i(t)$ encodes the body's global-frame center-of-mass worldline, where
\be\label{zcm}
z_A^i(t)=z_A^i(s_A)\Big|_{s_A=s^0_A(t)}
\ee
is the quantity $z^A_i(s_A)$ expressed as a function of $t$, with the function $s^0_A(t)$ found by setting $y_A^i=0$ in Eq.~(\ref{t}) (see Eq.~(\ref{sA0}) and discussion thereabouts).  The translational equation of motion for a body $A$, discussed in the next subsection, can be written in the form of a second-order ODE for the global-frame CoM worldline $z_A^i(t)$.

\subsection{Single-body laws of motion and translational equations of motion}\label{loms}

The single-body laws of motion are constraints on the lowest-order multipole moments of any body which govern the exchange of energy, momentum, and angular momentum between the body and the gravitational field.  The laws of motion at 1PN order were first found by DSX, who derived them by using covariant stress-energy conservation at 1PN order in the interior of the body.  The same laws of motion were later rederived by RF by using the 2PN (next-to-next-to-leading order in $1/c^2$) vacuum Einstein equation in a buffer region surrounding the body, thus extending their range of validity to include bodies with strong internal gravity.

The laws of motion are written in terms of the body's multipole and tidal moments as defined by the expansion of the 1PN potentials (\ref{potA}) and are valid in any coordinate system in which the spacetime metric takes the form given by (\ref{metric}) and (\ref{potA})---not just in body-adapted coordinates.  The results are
\bse\label{lom_}
\begin{widetext}
\bea
\dot{M}_A &=& - \frac{1}{c^2}\sum_{l=0}^\infty \frac{1}{l!} \left[ (l+1) \n M_A^L
  \n \dot{G}_A^L + l \n\dot{M}_A^L \n G_A^L \right]+O(c^{-4}),                          \label{lom_m}
\\
\ddot{M}_A^i&=&
\sum_{l=0}^\infty\frac{1}{l!}\bigg\{
M_A^L G_A^L
+\frac{1}{c^2}\bigg[
  \frac{1}{l+2}\epsilon_{ijk}M_A^{jL}\dot{H}_A^{kL} +
  \frac{1}{l+1}\epsilon_{ijk}\dot{M}_A^{jL}H_A^{kL}
 \nnm \\  & & 
  -\frac{2l^3 +7l^2 +15l +
    6}{(l+1)(2l+3)}M_A^{iL}\ddot{G}_A^L - \frac{2l^3 + 5l^2 +12l +
    5}{(l+1)^2}\dot{M}_A^{iL}\dot{G}_A^L - \frac{l^2 + l +
    4}{l+1}\ddot{M}_A^{iL}G_A^L
    \nnm \\ & & 
+ \frac{l}{l+1}S_A^L H_A^{iL}
- \frac{4(l+1)}{(l+2)^2}\epsilon_{ijk}S_A^{jL}\dot{G}_A^{kL}
- \frac{4}{l+2}\epsilon_{ijk}\dot{S}_A^{jL}G_A^{kL}
  \bigg]\bigg\}+O(c^{-4}).                                                              \label{lom_mi}
\\
{\dot S}_A^i &=& \sum_{l=0}^\infty \frac{1}{l!} \epsilon_{ijk} M_A^{jL} G_A^{kL}+O(c^{-2}),   \label{lom_s}
\eea
\end{widetext}
\ese
Eq.~(\ref{lom_m}) shows that the mass monopole $M_A$ is conserved at Newtonian order ($O(c^0)$), but not at 1PN order.  As discussed further in Sec.~\ref{truncate} below, $M_A$ contains $O(c^{-2})$ contributions from the internal energy of the body, which can vary as tidal forces do work on the body.
The law of motion (\ref{lom_s}) for the the spin $S_A^i$ is the same Newtonian-order tidal torque formula found in Eq.~(\ref{nspineom}).

The law of motion (\ref{lom_mi}) for the mass dipole $M_A^i$ governs the evolution of the body's total linear momentum.  The body's translational equation of motion can be derived by applying (\ref{lom_mi}) in the body frame, i.e.~by applying it to the body-frame dipole moment, as follows.

At Newtonian order, the $O(c^0)$ part of $\ddot{M}_A^i$ in (\ref{lom_mi}) gives the net force acting on the body, as $\dot M_A^i$ is the body's total momentum (cf. (\ref{siM})).  Since the body-adapted coordinates are chosen to be mass-centered $(M_A^i=0)$, this net force must vanish in the body frame.  This apparent equilibrium in the body frame is achieved by the balancing of gravitational forces from the other bodies with \emph{inertial} forces, which are due to the fact that the body frame is accelerating with respect to the (asymptotically) inertial global frame, along the worldline $z_A^i(t)$.  Both of these effects are accounted for by the body-frame tidal moments $G_A^L$; from the $O(c^0)$ part of [ref], we have
\bea
G_A^i&=&G_\gA^i-\ddot{z}_A^i+O(c^{-2}), \nnm
\\
G_A^L&=&G_\gA^L+O(c^{-2}),\quad(l\ge2), \nnm
\\
G_\gA^L&=&\sum_{B\ne A}\sum_{k=0}^\infty\frac{(-1)^k}{k!}M_B^K\doe^{(A)}_{KL}\frac{1}{|\bm{z}_A-\bm{z}_B|}+O(c^{-2}) \nnm
\eea
Using these relations in (\ref{lom_mi}), the requirement of equilibrium in the body frame, $\ddot M_A^i=0$, determines the equation of motion for the worldline $z_A^i(t)$:
\bea\label{newteom2}
&&M^A \ddot{z}^A_i=\sum_{B\ne A}\sum_{l=0}^\infty \frac{1}{l!}M^A_L G^{g,A}_{iL}+O(c^{-2})
\\ \nnm
&&=\sum_{B\ne A}\sum_{k,l=0}^\infty \frac{1}{k!l!}M^A_L M^B_K \, \doe^{(A)}_{iKL}\frac{1}{|\bm{z}^A-\bm{z}^B|}+O(c^{-2}).
\eea
This matches the Newtonian equation of motion found above in (\ref{genneom}).

At 1PN order, the procedure is essentially the same, but more involved.  One begins by setting $\ddot M_A^i=0$ in the body frame, with $\ddot M_A^i$ given by (\ref{lom_mi}).  To arrive at a suitable form for the final equation of motion, one must then rewrite the body-frame tidal moments $G_A^L$ and $H_A^L$ of body $A$ in terms of the body-frame multipole moments $M_B^L$ and $S_B^L$ of the other bodies $B\ne A$ and the worldline data $\mc{D}_C$ for all the bodies $C$; the details of this procedure are presented in Appendix \ref{atm}.

In the end, one arrives at an expression for the acceleration $\ddot z_A^i(t)$ of the 1PN-accurate global-frame center-of-mass worldline $z_A^i(t)$, defined by (\ref{zcm}).  (As in the Newtonian case, the acceleration term, describing inertial forces in the body frame, emerges from the transformation laws for the body-frame tidal moments.)  The expression depends only on the body-frame mass and current multipole moments $M_B^L(t)$ and $S_B^L(t)$,\footnote{
Here, and throughout, $M_A^L(t)$ and $S_A^L(t)$ are the body-frame moments $M_A^L(s_A)$ and $S_A^L(s_A)$ expressed as functions of $t$ at $y_A^i=0$, i.e.~the same physical quantities expressed as functions of different variables; cf.~Eq.~(\ref{sA0}) and surrounding discussion.}
the global-frame worldlines $z_B^i(t)$, and the time derivatives of these quantities, for all bodies $B$:
\bse\label{forms}
\be\label{zddform}
\ddot z_A^i(t)=\mc{F}_A^i(z_B^i,\dot z_B^i,M_B^L,\dot{M}_B^L,\ddot{M}_B^L,S_B^L,\dot S_B^L).
\ee
Similar (though simpler) manipulations applied to the laws of motion (\ref{lom_m}) and (\ref{lom_s}) allow one to write equations of motion for the mass monopole $M_A(t)$ and spin $S_A^i(t)$ in terms of the same variables:
\bea\label{mdform}
\dot{ M}_A(t)&=&\mc F_A(z_B^i,\dot z_B^i,{M}_B^L,\dot{{M}}_B^L),
\\
\dot S_A^i(t)&=&\tilde{\mc F}_A^i(z_B^i,{M}_B^L).\label{sdform}
\eea
\ese
The explicit forms of the translational equations of motion (\ref{zddform}) and the mass and spin evolution equations (\ref{mdform}) and (\ref{sdform}) will be given for the $M_1$-$M_2$-$S_2$-$Q_2$ case in Sec.~\ref{truncate}, and are given in the fully general case in RF \cite{RF} as corrected by an upcoming erratum.

To arrive at a closed set of evolution equations for the quantities $z_A^i(t)$, $M_A^L(t)$, and $S_A^L(t)$, for all bodies $A$, the equations of motion (\ref{forms}) must be supplemented by equations for the multipole moments $M^A_L(t)$ and $S^A_L(t)$ for $l\ge 2$.  Finding such equations will require a model for the bodies' internal dynamics, which will be addressed in Sec.~\ref{adiabatic}.

\subsection{$M_1$-$M_2$-$S_2$-$Q_2$ truncation}\label{truncate}

We have presented above the formalism for treating the 1PN dynamics of a collection of many bodies, each with arbitrarily high-order multipole moments.  Here, we apply that formalism to the two-body system discussed in Sec.~\ref{results}, with a body 1 having only a mass monopole moment $M_1$, and a body 2 having a mass monopole $M_2$, a current dipole, or spin, $S_2^i\equiv S^i$, and a mass quadrupole $M_2^{ij}\equiv Q_2^{ij}\equiv Q^{ij}$.   More precisely, we truncate the internal parts of body-frame multipole series (\ref{potA}) for each body according to
\bea
\Phi_{1,{\rm{int}}}&=&-\frac{M_1}{|\bm y_1|}+O(c^{-4}),
\nnm\\
\zeta^i_{1,{\rm{int}}}&=&O(c^{-2}),
\nnm
\eea
neglecting the moments $M_1^L$ for $l\ge2$ and $S_1^L$ for $l\ge1$, and
\bea
\Phi_{2,{\rm{int}}}&=&-\frac{M_2}{|\bm y_2|}-\frac{1}{2}Q^{ij}\doe_{ij}\frac{1}{|\bm y_2|}
\nnm\\
&&-\frac{1}{4c^2}\ddot{Q}^{ij}\doe_{ij}|\bm y_2|+O(c^{-4}),
\nnm\\
\zeta^i_{2,{\rm{int}}}&=&2\left(\dot Q^{ij}-\epsilon_{ijk}S^k\right)\doe_j\frac{1}{|\bm y_2|}+O(c^{-2}),
\nnm
\eea
neglecting the moments $M_2^L$ for $l\ge3$ and $S_2^L$ for $l\ge2$.  (The external parts of these potentials will be just as in (\ref{potA}), with arbitrarily higher-order tidal moments.)  These expressions for the body-frame potentials define the body-frame moments $M_1(s_1)$, $M_2(s_2)$, $Q^{ij}(s_2)$, and $S^i(s_2)$ as functions of the body-frame time coordinates $s_1$ and $s_2$.  These moments can be expressed as functions of the global time coordinate, written $M_1(t)$, $M_2(t)$, $S^i(t)$ and $Q^{ij}(t)$, using the coordinate transformation (\ref{xt}) with $y_A^i=0$ (cf. Eq.~(\ref{sA0})).  For the bodies' global-frame CoM worldlines $z_1^i(t)$ and $z_2^i(t)$, we will use the definitions
\be\label{rel}
z^i=z_2^i-z_1^i,\quad r=|\bm z|,\quad n^i=z^i/r,
\ee
as similar to (\ref{reln}), except that the worldlines are now defined with 1PN accuracy, and
\be
v_1^i=\dot z_1^i,\quad v_2^i=\dot z_2^i,\quad v^i=v_2^i-v_1^i.
\ee

With these conventions in place, we can apply the laws of motion presented in Sec.~\ref{loms} to find the evolution equations for the moments $M_1(t)$, $M_2(t)$, and $S^i(t)$ and the global-frame center-of-mass worldlines $z_1^i(t)$ and $z_2^i(t)$.  The results involve only these quantities and the quadrupole $Q^{ij}(t)$.

As body 1 has no higher-order multipole moments, the law of motion (\ref{lom_m}) requires that its mass monopole $M_1$ be constant in time:
\be\label{M1dot}
\dot M_1=O(c^{-4}).
\ee
The same law of motion applied to body 2 gives
\be\label{M2dot}
\dot{M}_2=-\frac{1}{c^2}\left(\frac{3}{2}Q^{ij}\dot{G}_2^{ij}-\dot{Q}^{ij}G_2^{ij}\right)+O(c^{-4}),
\ee
where the body-frame gravito-electric tidal moment $G_2^{ij}$ is given by (\ref{FgA}) as
\be\label{nG2ij}
G_2^{ij}=G_{{\rm{g}},2}^{ij}+O(c^{-2})=\frac{3M^A}{r^3}n^{<ij>}+O(c^{-2}).
\ee
It is worth pausing here to compare this rate of change of the 1PN-accurate mass monopole $M_2$ with the rate of change Newtonian internal energy $E_2^{\rm{int}}$ discussed in Sec.~\ref{nEnergy}.  From (\ref{E2intdot}) and (\ref{M2dot}), we find that they are related by
\be
\dot M_2=c^{-2}\left(\dot E_2^{\rm{int}}+3\dot U_Q\right)+O(c^{-4}),
\ee
where $U_Q$ is the Newtonian gravitational potential energy associated with the quadrupole-tidal interaction given by (\ref{UQ}).  The mass monopole $M_2$ thus contains contributions not only from body 2's internal energy but also from the tidal part of its external gravitational potential energy.  Though Newtonian internal energy is not a well-defined concept for strongly self-gravitating bodies, we will simply take the relation
\be\label{M2split}
M_2=\n M_2 + c^{-2}\left(E_2^{\rm{int}}+3 U_Q\right)+O(c^{-4})
\ee
to define the quantity $E_2^{\rm{int}}$ in the 1PN context, with $\n M_2$ being the conserved Newtonian-order rest-mass contribution.
  This partitioning of $M_2$, which is fully consistent with the equation of motion (\ref{M2dot}), given (\ref{E2intdot}) and (\ref{UQ}), will be useful in our discussion of internal dynamics below.  (A thorough discussion of the ambiguity in the total mass-energy of a body by an amount of the order of its tidal potential energy, and of tidal heating in GR, can be found in Ref.~\cite{Purdue}.)

The evolution equation for the spin (\ref{lom_s}) gives the following tidal torque on body 2:
\be\label{Sdot}
\dot S^i= \epsilon^{ijk} Q^{aj} G_2^{ka}+O(c^{-2}),
\ee
with $G_2^{ka}$ being given by (\ref{nG2ij}).  This coupling between the spin and the quadrupole, which is a purely Newtonian effect, is the essential reason we cannot (in general) ignore the spin-orbit coupling terms in the 1PN translational equations of motion for bodies with quadrupole moments.

Finally, working from the laws of motion (\ref{lom_mi}) for the mass dipoles, we can apply the procedure outlined in Sec.~\ref{loms} and Appendix \ref{atm} to the $M_1$-$M_2$-$S_2$-$Q_2$ system to find the translational equations of motion for the worldlines $z_1^i$ and $z_2^i$.  The results are
\bse\label{teoms}
\bea
M_1\ddot z_1^i(t)=F_{1,M}^i+F_{1,S}^i+F_{1,Q}^i,\label{zdd1}
\\
M_2\ddot z_2^i(t)=F_{2,M}^i+F_{2,S}^i+F_{2,Q}^i,\label{zdd2}
\eea
with the monopole contributions,
\begin{widetext}
\bea
F_{1,M}^i&=&\frac{M_2}{r^2}n^i+\frac{1}{c^2}\frac{M_2}{r^2}\left\{n^i\left[2v^2-v_1^2-\frac{3}{2}(n^av_2^a)^2-\frac{5M_1}{r}-\frac{4M_2}{r}\right]
+v^in^a(4v_1^a-3v_2^a)\right\}+O(c^{-4}),\label{F1M}
\\
F_{2,M}^i&=&-\frac{M_1}{r^2}n^i-\frac{1}{c^2}\frac{M_1}{r^2}\left\{n^i\left[2v^2-v_2^2-\frac{3}{2}(n^av_1^a)^2-\frac{4M_1}{r}-\frac{5M_2}{r}\right]
+v^in^a(4v_2^a-3v_1^a)\right\}+O(c^{-4}),\label{F2M}
\eea
the spin contributions,
\bea
F_{1,S}^i&=&\frac{1}{c^2}\frac{M_1}{r^3}\epsilon^{abc}S^c\Big[\delta^{ai}(4v^b-6n^{bd}v^d)-6n^{ai}v^b\Big]+O(c^{-4}),\label{F1S}
\\
F_{2,S}^i&=&\frac{1}{c^2}\frac{M_1}{r^3}\epsilon^{abc}S^c\Big[3\delta^{ai}(n^{bd}v^d-v^b)+6n^{ai}v^b\Big]+O(c^{-4}),\label{F2S}
\eea
and the quadrupole contributions,
\bea
F_{1,Q}^i&=&\frac{3M_1}{2r^4}Q^{ab}\left(5n^{abi}-2n^a\delta^{bi}\right)
+\frac{1}{c^2}\Bigg(
\frac{3M_1}{2r^4}Q^{ab}\bigg\{5n^{abi}\left[2v^2-v_1^2-\frac{7}{2}(n^cv_2^c)^2-\frac{47M_1}{5r}-\frac{24M_2}{5r}\right]
\nnm\\
&&-2n^a\delta^{bi}\left[2v^2-v_1^2-\frac{5}{2}(n^cv_2^c)^2-\frac{19M_1}{2r}-\frac{4M_2}{r}\right]
+n^av_2^{bi}+(5n^{ai}-\delta^{ai})v_2^{bc}n^c
\nnm\\
&&+v^i(5n^{abc}-2n^a\delta^{bc})(4v_1^c-3v_2^c)\bigg\}
+\frac{3M_1}{2r^3}\dot Q^{ab}\Big[n^{ab}(5v_2^{c}n^{ci}+3v^i)-4v^an^{bi}-2\delta^{ai}n^{bc}(2v_1^c-v_2^c)\Big]
\nnm\\
&&-\frac{3M_1}{4r^2}\ddot Q^{ab}\big(n^{abi}+2n^a\delta^{bi}\big)\Bigg)+O(c^{-4}),\label{F1Q}
\\
F_{2,Q}^i&=&-\frac{3M_1}{2r^4}Q^{ab}\left(5n^{abi}-2n^a\delta^{bi}\right)
+\frac{1}{c^2}\Bigg(
\frac{3M_1}{2r^4}Q^{ab}\bigg\{-5n^{abi}\left[2v^2-v_2^2-\frac{7}{2}(n^cv_1^c)^2-\frac{8M_1}{r}-\frac{6M_2}{r}\right]
\nnm\\
&&+2n^a\delta^{bi}\left[3v^2-v_2^2-5(n^cv^c)^2-\frac{5}{2}(n^cv_1^c)^2-\frac{8M_1}{r}-\frac{11M_2}{2r}\right]
+n^iv^{ab}+5n^{aci}(2v^bv_1^c-v_2^{bc})
\nnm\\
&&+v^i(5n^{abc}-2n^a\delta^{bc})(4v_2^c-3v_1^c)+n^av_2^b(v_2^i-2v_1^i)
+\delta^{bi}n^c\left[(5v_2^a-4v_1^a)v_2^c-6v^av_1^c\right]\bigg\}
\nnm\\
&&+\frac{3M_1}{r^3}\dot Q^{ab}\Big[v^b(2n^{ai}-\delta^{ai})+\delta^{ai}n^{bc}v^c-2n^{ab}v^i\Big]\Bigg)+O(c^{-4}).\label{F2Q}
\eea
\end{widetext}
\ese
(It should be noted that occurrences of $\dot S^i$ in the equations of motion have been replaced by (\ref{Sdot}) and included in the quadrupole contributions.)  

The monopole contributions (\ref{F1M},\ref{F2M}) give the well-known Lorentz-Droste-Einstein-Infeld-Hoffmann accelerations, and the spin contributions (\ref{F1S},\ref{F2S}) give the well-known 1PN spin-orbit terms \cite{DSX2}.  The quadrupole contributions (\ref{F1Q},\ref{F2Q}) have been derived previously by Xu, Wu, and Schafer \cite{XuWuSchafer}, though our results disagree with theirs in several terms; we have not been able to pin down the source of the disagreement.  Our results also disagree with the final results of RF \cite{RF}, but agree with their corrected results given in an upcoming erratum.  The strongest indication of the correctness of our expressions for the EoMs is the fact that, unlike the results in \cite{XuWuSchafer,RF}, they are consistent with the conservation of the binary system's total linear momentum, as discussed in Sec.~\ref{sysdipole} below.  We can also note that the action derived below from these results agrees with the recent work (using a rather different method) by Damour and Nagar \cite{DN2} and Bini, Damour, and Faye \cite{BDF}.

\section{System multipole moments and conservation laws}\label{system}

In Sec.~\ref{body}, we defined the multipole moments of a single body through the multipole expansion of the metric in a vacuum buffer region surrounding the body.  The same procedure can be applied to a collection of several bodies to define multipole moments for the entire system.  Applying the general laws of motion discussed in Sec.~\ref{loms} to these system multipole moments will allow us to formulate conservation laws for the energy, momentum, and angular momentum of an isolated $N$-body system, expressed as constraints on the worldlines and multipole moments of the constituent bodies.  These conservation laws can serve both as a consistency check for the equations of motion given in Sec.~\ref{truncate} and as a means to specialize the equations of motion to the system's center-of-mass frame.

\subsection{General formulae}\label{genform}

We have already discussed, in Sec.~\ref{global}, a form for the metric generated by a system of $N$ bodies.  Using the global coordinate system $(t,x^i)$, we expressed the potentials parameterizing the metric as a sum of multipole expansions for each body $A$, written in terms of the bodies' global-frame multipole moments $M_\gA^L$ and $Z_\gA^{iL}$ and Newtonian-order worldlines $z_A^i$:
\begin{widetext}
\bse\label{potgC}
\bea
\Phi_{\rm{g}}&=&-\sum_{A}\sum_{l=0}^{\infty}\frac{(-1)^{l}}{l!}\left[M_\gA^L \doe_L\frac{1}{|\bm x-\bm z_A|}
+\frac{1}{2c^2}\doe_t^2\left(M_\gA^L \doe_L|\bm x-\bm z_A|\right)\right]+O(c^{-4}),\label{phigC}
\\
\zeta_{\rm{g}}^i&=&-\sum_{A}\sum_{l=0}^{\infty}\frac{(-1)^{l}}{l!}Z_{{\rm{g}},A}^{iL} \doe_L\frac{1}{|\bm x-\bm z_A|}+O(c^{-2}),\label{zetagC}
\eea
\ese
\end{widetext}
with $Z_\gA^{iL}$ satisfying (\ref{Zgconds}).  This solution for the metric was constructed to be valid in the region $\mc B_{\rm{g}}$, which extends out to spatial infinity.

In a region far outside the system, we can rewrite these expressions for the global-frame potentials to mirror the forms (\ref{potA}) used to define the multipole moments of a single body, with multipole expansions about the global-frame origin $x^i=0$:
\begin{widetext}
\bse\label{potgsys}
\bea
\Phi_{\rm{g}}&=&-\sum_{l=0}^{\infty}\frac{(-1)^{l}}{l!}\left\{M_{\rm{sys}}^L\doe_L\frac{1}{|\bm x|}
+\frac{1}{c^2}\left[\frac{(2l+1)}{(l+1)(2l+3)}\dot{\mu}_{\rm{sys}}^L\doe_L\frac{1}{|\bm x|}
+\frac{1}{2}\n \ddot M_{\rm{sys}}^L \doe_L|\bm x|\right]\right\}+O(c^{-4}),\label{phigsys}
\\
\zeta_{\rm{g}}^i&=&-\sum_{l=0}^{\infty}\frac{(-1)^{l}}{l!}Z_{\rm{sys}}^{iL} \doe_L\frac{1}{|\bm x|}+O(c^{-2}),\label{zetagsys}
\eea
\ese
with
\be
Z_{\rm{sys}}^{iL}=\frac{4}{l+1}\n\dot{M}_{\rm{sys}}^{iL}-\frac{4l}{l+1}\epsilon^{ji<a_l}S_{\rm{sys}}^{L-1>j}
+\frac{2l-1}{2l+1}\delta^{i<a_l}\mu_{\rm{sys}}^{L-1>}+O(c^{-2}).
\ee
\end{widetext}
These expansions define the system multipole moments $M_{\rm{sys}}^L$ and $S_{\rm{sys}}^L$ and the gauge moments $\mu_{\rm{sys}}^L$.  The tidal terms present in (\ref{potA}) are absent here, as the global-frame potentials vanish as $|\bm x|\to\infty$ (cf.~Eq.~(\ref{potgC})).  To compare the metric (\ref{potgsys}) here to the metric (\ref{potgC}) above, we must express them in the same gauge.  We have chosen the gauge that enforces $Z_{{\rm{g}},A}^{jjL}=0$ in (\ref{potgC}), which will result in nonzero values for the system gauge moments $\mu_{\rm{sys}}^L$ in (\ref{potgsys}).

Since the potentials given by (\ref{potgC}) and by (\ref{potgsys}) represent the same metric in the same gauge, they should be explicitly equal.  This condition will allow us to solve for the system multipole moments appearing in (\ref{potgsys}) in terms of the individual bodies' global-frame multipole moments and worldlines appearing in (\ref{potgC}).

Considering first the vector potential $\zeta^i_{\rm{g}}$, we can use the Taylor series
\be\label{taylor}
|\bm x-\bm z_A|^n=\sum_{k=0}^\infty\frac{(-1)^k}{k!} z_A^K \doe_K |\bm x|^n
\ee
to rewrite (\ref{zetagC}) in the form
\bea
&&\zeta^i_{\rm{g}}=-\sum_A\sum_{l,k=0}^\infty\frac{(-1)^{l+k}}{l!k!}Z_\gA^{iL} z_A^K \doe_{LK} \frac{1}{|\bm x|}
\nnm\\ \nnm
&&=-\sum_A\sum_{p=0}^\infty\sum_{k=0}^p\frac{(-1)^{p}}{p!}\frac{p!}{k!(p-k)!}Z_\gA^{<P-K} z_A^{K>}\doe_P\frac{1}{|\bm x|}.
\eea
In the second line, we have relabeled the multi-indices according to $LK\to P$, adjusted the summations accordingly, and used the fact that $\doe_P|x|^{-1}$ is STF.  Renaming $P\to L$ and comparing this with (\ref{zetagsys}) gives an expression for the tensors $Z_{\rm{sys}}^{iL}$:
\bse\label{gravmagsys}
\be\label{ZiLsys}
Z_{\rm{sys}}^{iL}=\sum_A\sum_{k=0}^l\frac{l!}{(l-k)!k!} Z_{{\rm{g}},A}^{i<L-K} z_A^{K>}.
\ee
The current moments and gauge moments can then be found from formulae analogous to (\ref{ZtoS}) and (\ref{Ztomu}):
\bea
S_{\rm{sys}}^L&=&\frac{1}{4}Z_{\rm{sys}}^{jk<L-1}\epsilon^{a_l>kj}
\\
\mu_{\rm{sys}}^L&=&Z_{\rm{sys}}^{jjL}
\eea
\ese

The scalar potential $\Phi_{\rm{g}}$ can be manipulated in a similar manner, using the Taylor series (\ref{taylor}), to find formulae for the system mass multipole moments $M_{\rm{sys}}^L$.  The details of this more involved procedure are given in Appendix \ref{sysmmm}.  The result is
\bea
&&M_{\rm{sys}}^L=\sum_A\sum_{k=0}^l\frac{l!}{k!(l-k)!}\Bigg[ M_\gA^{<L-K} z_A^{K>}
\nnm\\
&&+\frac{1}{c^2}\frac{1}{2(2l+3)}\doe_t^2\left(2 M_\gA^{j<L-K} z_A^{K>j}+ M_\gA^{<L-K} z_A^{K>jj}\right)\Bigg]
\nnm\\
&&-\frac{1}{c^2}\frac{2l+1}{(l+1)(2l+3)}\dot{\mu}_{\rm{sys}}^L+O(c^{-4}).
\label{MLsys}
\eea
The gauge moments $\dot{\mu}_{\rm{sys}}^L$ appearing here can be found from (\ref{gravmagsys}).

In summary, Eqs.~(\ref{gravmagsys}) and (\ref{MLsys}) give the total multipole moments $M_{\rm{sys}}^L$ and $S_{\rm{sys}}^{L}$ of an $N$-body system, defined in the global frame $(t,x^i)$, in terms of the individual bodies' global-frame multipole moments $M_{{\rm{g}},A}^L$ and $Z_{{\rm{g}},A}^{iL}$ and worldlines $z_A^i$.  In the following subsections, considering the two-body $M_1$-$M_2$-$S_2$-$Q_2$ case, we will use these results along with the moment transformation formulae from Appendix \ref{atm} to write the system's mass monopole $M_{\rm{sys}}$ and mass dipole $M_{\rm{sys}}^i$ in terms of the body-frame moments $(M_1,M_2,S^i,Q^{ij})$ and the worldlines $z_1^i$ and $z_2^i$.  We note that a similar procedure can be applied to find $S_{\rm{sys}}^i=\epsilon^{ijk}(M_1z_1^jv_1^k+M_2z_2^jv_2^k)+S_2^i$ for the system's total (Newtonian) angular momentum.  The system's 1PN accurate mass quadrupole, which will be needed for the calculation of the gravitational wave signal from the binary system, can also be calculated from Eq.~(\ref{MLsys}).

\subsection{System mass monopole}\label{sysmonopole}

Specializing the general formula (\ref{MLsys}) for the system mass multipoles to the monopole $(l=0)$ case, and using the $M_1$-$M_2$-$S_2$-$Q_2$ truncation, we find the binary system's total 1PN-accurate mass monopole to be
\bea
M_{\rm{sys}}&=&M_{{\rm{g}},1}+M_{{\rm{g}},2}+\frac{1}{c^2}\bigg[-\frac{1}{3}\dot\mu_{\rm{sys}}
\nnm\\
&&+\frac{1}{6}\frac{d^2}{dt^2}\left(M_{{\rm{g}},1} z_1^2+ M_{{\rm{g}},2} z_2^2\right)\bigg]+O(c^{-4}).
\nnm
\eea
Using (\ref{gravmagsys}) for $\dot\mu_{\rm{sys}}$ and the formulae in Appendix \ref{bfmult_gfmult} relating the global- and body-frame multipole moments, we can rewrite this expression in terms of the body-frame multipole moments and the CoM worldlines:
\bea
M_{\rm{sys}}&=&M_1+M_2+\frac{1}{c^2}\bigg(\frac{M_1v_1^2}{2}+\frac{M_2v_2^2}{2}
\nnm\\ \nnm
&&-\frac{M_1M_2}{r}-2U_Q\bigg)+O(c^{-4}),
\eea
where the tidal potential energy $U_Q$, as in (\ref{UQ}), is
\bdm
U_Q=-\frac{3M_1}{2r^3}Q^{ij}n^{ij}.
\edm
If we rewrite the mass monopole of body 2 as $M_2=\n M_2 + c^{-2}\left(E_2^{\rm{int}}+3 U_Q\right)$, as in (\ref{M2split}), we find that the 1PN contribution to the system mass monopole is exactly the system's total Newtonian energy $E$ given by (\ref{nE}):
\bea
M_{\rm{sys}}&=&M_1+\n M_2+c^{-2}E+O(c^{-4})
\\
E&=&\frac{1}{2}M_1v_1^2+\frac{1}{2}M_2v_2^2-\frac{M_1M_2}{r}+U_Q+E_2^{\rm{int}}. \nnm
\eea
This is a further validation of the decomposition of the total mass monopole $M_2$ in Eq.~(\ref{M2split}).  The constancy of $M_{\rm{sys}}$, required by the law of motion (\ref{lom_m}) as applied to the entire system in the global frame (for which there are no tidal moments), then follows from the constancy of $E$.

\subsection{System mass dipole}\label{sysdipole}

Taking the $l=1$ case in the general formula (\ref{MLsys}) gives the system's 1PN-accurate mass dipole:
\bea
M^i_{\rm{sys}}=M_\go z_1^i+M_\gt z_2^i +M_\go^i+M_\gt^i +\frac{1}{c^2}\bigg[-\frac{3}{10}\dot{\mu}_{\rm{sys}}^{i} &&
\nnm\\ \nnm
+\frac{1}{10}\doe_t^2\left(2 M_{{\rm{g}},2}^{ij} z_2^{j}+ M_\go z_1^{ijj}+ M_\gt z_2^{ijj}\right)\bigg]+O(c^{-4}) &&
\eea
Using (\ref{gravmagsys}) for $\dot{\mu}_{\rm{sys}}^{i}$, (\ref{Sdot}) to replace an occurrence of $\dot S^i$, (\ref{M2split}) to replace $M_2$, (\ref{UQ}) for $U_Q$, and the moment transformations from Appendix \ref{bfmult_gfmult}, this becomes
\begin{widetext}
\bea\label{Misys}
M^i_{\rm{sys}}&=&M_1z_1^i+\n M_2z_2^i+\frac{1}{c^2}\Bigg[z_1^i\left(\frac{M_1v_1^2}{2}-\frac{M_1M_2}{2r}+\frac{U_Q}{2}\right)
\nnm\\
&&+z_2^i\left(\frac{M_2v_2^2}{2}-\frac{M_1M_2}{2r}+\frac{U_Q}{2}+E_2^{\rm{int}}\right)
+\frac{3M_1}{2r^2}Q^{ij}n^j  + \epsilon^{ijk} v_2^j S^{k} \Bigg]+O(c^{-4})
\eea
\end{widetext}
From the law of motion (\ref{lom_mi}) as applied to the entire system in the global frame, for which there are no tidal moments, we see that $\ddot M^i_{\rm{sys}}$ should vanish; this is a statement of total momentum conservation.  By differentiating (\ref{Misys}), order reducing as appropriate, using the full 1PN translational equations of motion (\ref{teoms}) in the Newtonian terms and their Newtonian parts in the 1PN terms, and also using (\ref{E2intdot}) and (\ref{Sdot}) for $\dot{E}_2^{\rm{int}}$ and $\dot{S}_{i}$, we find that indeed $\ddot M^i_{\rm{sys}}=0$.  This is an important check of the correctness of our expressions for the equations of motion and of the consistency of the formalism (which is not satisfied by the EoMs given in Ref.~\cite{RF,XuWuSchafer}).

Note that the inclusion of the spin term in Eq.~(\ref{Misys}) is essential for this consistency check, reflecting the necessity of including spin terms when working with generic mass quadrupoles at 1PN order.  Anecdotally, at the beginning of this work, we tried working through the DSX formalism dropping all spin terms while keeping quadrupoles, and we obtained certain EoMs. ÊTo check our results, we wanted to show that the momentum given by the time derivative of Eq. (4.8) (without the final term, the spin term) vanished as a result of our EoMs, which it did not. ÊOnly with the inclusion of the spin term in (4.8) and with all spin terms in the EoMs were we able to ensure momentum conservation for completely generic evolution of the quadrupoles.

\section{Orbital dynamics in the system's center-of-mass frame}\label{dynamics}

\subsection{Equation of motion of the relative position}\label{comframe}

The conservation of momentum allows us to reduce the problem of solving for the two worldlines $z_1^i(t)$ and $z_2^i(t)$ to solving for just their separation $z^i(t)=z_2^i(t)-z_1^i(t)$ in the binary system's center-of-mass (CoM) frame.  We can define the CoM frame to be that in which the 1PN-accurate mass dipole vanishes,
\be
M_{\rm sys}^i(t)=0,
\ee
so that the system's center-of-mass(-energy) is at rest at the global-frame spatial origin.  This fixes all remaining (post-Galilean) coordinate freedom in the global-frame metric.   Using Eq.~(\ref{Misys}), this condition can be used to solve for the worldlines $z_1^i$ and $z_2^i$ in the global CoM frame in terms of the relative position $z^i$ (working perturbatively in $c^{-2}$); one finds
\bea\label{ztoz1}
z_1^i&=&-\chi_2z^i+c^{-2}\left(\mathcal{P}z^i-\mathcal{D}^i\right)+O(c^{-4}),
\\
z_2^i&=&\chi_1z^i+c^{-2}\left(\mathcal{P}z^i-\mathcal{D}^i\right)+O(c^{-4}),\label{ztoz2}
\eea
with
\bea
\mathcal{P}&=&\eta(\chi_2-\chi_1)\left(\frac{v^2}{2}-\frac{M}{2r}-\frac{3}{4\chi_2 r^3}Q^{ij}n^{ij}\right)
-\frac{\chi_1}{M}E_2^{\rm{int}},
\nnm\\
\mathcal{D}_i&=&\frac{3\chi_i}{2r^2}Q^{ij}n_j+\frac{\chi_1}{M}\epsilon^{ijk}v^jS^k,
\eea
and with the new notation
\bea
&M=M_1+\n M_2,\quad \chi_1=M_1/M,\quad \chi_2=\n M_2/M,&
\nnm\\
&\mu=M_1 \n M_2/M,\quad \eta=\chi_1\chi_2=\mu/M.&
\eea

To find the acceleration of the relative position in the CoM frame, we can simply subtract our above results (\ref{teoms}) for the individual accelerations:
\be
a^i\equiv\ddot{z}^i=\ddot{z}_2^i-\ddot{z}_1^i.
\ee
The resulting expression depends only on $z^i$, $v_1^i$, $v_2^i$, $M_1$, $M_2$, $Q^{ij}$, $S^i$, and $E_2^{\rm{int}}$.  As $v_1^i$ and $v_2^i$ appear only in 1PN terms, we can replace them with their Newtonian values in the CoM frame,
\be\label{comvs}
v_1^i=-\chi_2 v^i+O(c^{-2}),\quad v_2^i=\chi_1 v^i+O(c^{-2}),
\ee
from differentiating the $O(c^0)$ parts of (\ref{ztoz1},\ref{ztoz2}).  After defining one last shorthand,
\be
\dot{r}=n^av^a,
\ee
we can write our result for the 1PN-accurate CoM-frame relative acceleration as follows:
\bse\label{aCoM}
\be
a^i=a^i_M+a^i_S+a^i_Q,
\ee
with the monopole contribution,
\begin{widetext}
\be\label{aM}
a^i_M=-\frac{M}{r^2}n^i  -\frac{1}{c^2}\frac{M}{r^2} \left\{n^i\left[(1+3\eta)v^2 - \frac{3\eta}{2}\dot{r}^2
- 2 (2+\eta)\frac{M}{r}\right]-2(2-\eta)\dot{r}v^i\right\}+O(c^{-4}),
\ee
the spin contribution,
\be\label{aS}
a^i_S= \frac{\epsilon^{abc}S^c}{c^2\chi_2 r^3}\left[(3+\chi_2) v^a \delta^{bi} - 3(1+\chi_2) \dot{r} n^a \delta^{bi} + 2 n^{ai} v^b \right]+O(c^{-4}),
\ee
and the quadrupole contribution,
\bea
a^i_Q&=& - \frac{3Q^{ab}}{2\chi_2 r^4}\left[5n^{abi}-2 n^a \delta^{bi} \right]
+\frac{1}{c^2}\Bigg\{ \frac{Q^{ab}}{r^4} \bigg[ n^{abi} \left( B_1 v^2 + B_2 \dot{r}^2 + B_3 \frac{M}{r} \right)
\nnm\\
&&+ n^a \delta^{bi} \left( B_4 v^2 + B_5 \dot{r}^2 + B_6 \frac{M}{r} \right)
+ B_7 \dot{r} n^{ab} v^i + B_8 n^a v^{bi} + B_9 \dot{r} n^{ai} v^b + B_{10} v^{ab} n^i + B_{11} \dot{r} v^a \delta^{bi} \bigg]  \label{aQ}
\nnm\\
&& + \frac{\dot{Q}^{ab}}{r^3} \left[ B_{12} n^{ab}v^i + B_{13} \dot{r} n^{abi} + B_{14} n^{ai} v^b 
+ B_{15} v^a \delta^{bi} + B_{16} \dot{r} n^a \delta^{bi} \right]
+ \frac{\ddot{Q}^{ab}}{r^2}\left[B_{17} n^{abi} + B_{18} n^a \delta^{bi} \right]
\nnm\\&&- B_{19} \frac{E_2^{\rm{int}}}{r^2}n^i \Bigg\} +O(c^{-4}),
\eea
with coefficients
\bdm
B_1=-\frac{15}{2\chi_2}(1+3\eta),\:
B_2=\frac{105\chi_1}{4},\:
B_3=\frac{12}{\chi_2}(5-2\chi_2^2),\:
B_4=\frac{3}{\chi_2}(2+2\chi_2-3\chi_2^2),
\edm
\bdm
B_5=-\frac{15}{2\chi_2}(2-\chi_2-\chi_2^2),\:
B_6=-\frac{3}{\chi_2}(8-\chi_2-3\chi_2^2),\:
B_7=\frac{15}{\chi_2}(2-\eta),\:
B_8=-\frac{3}{2\chi_2}(7-2\chi_2+3\chi_2^2),
\edm
\bdm
B_9=-\frac{15\chi_1}{2\chi_2}(1+\chi_2),\:
B_{10}=\frac{3\chi_1}{2\chi_2},\:
B_{11}=\frac{3}{2\chi_2}(5-4\chi_2-\chi_2^2),\:
B_{12}=-\frac{3}{2\chi_2}(4-\chi_2),\:
B_{13}=-\frac{15\chi_1}{2},
\edm
\be\label{Bs}
B_{14}=\frac{6}{\chi_2},\:
B_{15}=-\frac{3\chi_1}{\chi_2},\:
B_{16}=\frac{3}{\chi_2}(1-2\chi_2-\chi_2^2),\:
B_{17}=\frac{3}{4},\:
B_{18}=\frac{3}{2},
B_{19}=1.
\ee
\end{widetext}
\ese
In this form for the CoM-frame orbital EoM, we have used (\ref{M2split}) to write the total 1PN-accurate mass monopole $M_2$ in terms of the (constant) Newtonian mass $\n M_2$, the internal energy $E_2^{\rm{int}}$, and the tidal potential energy $U_Q$ (giving a contribution to $B_3$).  This decomposition is useful in formulating an action principle for the orbital dynamics (as in the next subsection), as $E_2^{\rm{int}}$ is independent of the orbital degrees of freedom, while $M_2$ is not.

\subsection{Generalized Lagrangian for the orbital dynamics}\label{lagrangian}

The monopole contributions (\ref{aM}) to the 1PN CoM-frame orbital EoMs are known to be derivable from the Lagrangian
\bse\label{pnL}
\bea  \label{LM}
\mc L_M&=&\frac{\mu v^2}{2} + \frac{\mu M}{r} 
+\frac{\mu}{c^2}\bigg[ \frac{1-3\eta}{8}v^4 
\\ \nnm
&&+ \frac{M}{2r} \left( (3+\eta)v^2 + \eta \dot{r}^2 - \frac{M}{r} \right)\bigg]+O(c^{-4}),
\eea
(see e.g.~\cite{Brumberg}).  The spin contributions (\ref{aS}) can also be derived from an action principle, but with a generalized Lagrangian (one depending not only on the relative position $z^i$ and velocity $v^i=\dot z^i$, but also on the acceleration $a^i=\ddot z^i$) given by adding
\be\label{LS}
\mc L_S=\frac{\chi_1}{c^2}\epsilon^{abc}S^av^b\left[\frac{2M}{r^2}n^c+\frac{\chi_1 }{2}a^c\right]+O(c^{-4})
\ee
to (\ref{LM}) (see e.g.~\cite{Lspinorbit}).  Applying the generalized Euler-Lagrange equation,
\be\label{EL}
\left(\frac{\doe}{\doe z^i}-\frac{d}{dt}\frac{\doe}{\doe v^i}+\frac{d^2}{dt^2}\frac{\doe}{\doe a^i}\right)\mc L=0,
\ee
to $\mc L=\mc L_M+\mc L_S$, and using $\dot M_1=\dot M_2=\dot S^i=0$ (which replaces (\ref{M1dot},\ref{M2dot},\ref{Sdot}) in the case with no quadrupole), one recovers the EoM $a^i=a_M^i+a_S^i$ from (\ref{aM},\ref{aS}).  

We have found that the quadrupole contributions to the orbital EoM (\ref{aQ}) can also be encoded in a generalized Lagrangian.  To determine the necessary additions to the Lagrangian, one can proceed by guesswork, using the known Newtonian Lagrangian (\ref{nL}), and writing down all possible 1PN-order scalars that can be formed from the relative position $z^i$ and velocity $v^i$, the total (Newtonian) mass $M$, and linear factors of the quadrupole $Q^{ij}$, its time derivative $\dot Q^{ij}$, and the internal energy $E_2^{\rm{int}}$; including dimensionless coefficients $A_1$--$A_9$ for each such term, we have
\begin{widetext}
\bea
\mc L_Q&=&\frac{3M_1}{2 r^3} Q^{ab} n^{ab} 
+ \frac{1}{c^2} \Bigg\{ \frac{M}{r^3}Q^{ab} \left[ n^{ab} \left( A_1 v^2 + A_2 \dot{r}^2 + A_3 \frac{M}{r} \right) + A_4 v^{ab} + A_5 \dot{r} n^a v^b \right]
\nnm\\
&& \phantom{\frac{3M_1}{2 r^3} Q^{ab} n^{ab} + \frac{1}{c^2} \Bigg\{} + \frac{M}{r^2}\dot{Q}^{ab}\left[A_6 n^a v^b + A_7 \dot{r} n^{ab} \right] 
 + E^{\rm{int}}_2\left[A_8 v^2 + A_9 \frac{M}{r} \right] \Bigg\}+O(c^{-4}).   \label{LQ}
\eea
\end{widetext}
Since the spin-orbit terms (\ref{LS}) require the acceleration $a^i$, one might expect that terms with factors of $a^i$ and also $\ddot Q^{ij}$ should be included here; we find, however, that such terms are not necessary.  The only further term allowed by general considerations but not included here is $E_2^{\rm{int}}\dot r^2$, as recovering the EoM (\ref{aCoM}) requires its coefficient to be zero.

By applying the Euler-Lagrange equation (\ref{EL}) to the generalized Lagrangian $\mc L=\mc L_M+\mc L_S+\mc L_Q$, using the evolution equations (\ref{Sdot}) and (\ref{E2intdot}) for time derivatives of $S^i$ and $E_2^{\rm{int}}$, one finds an EoM of the same form as (\ref{aCoM}) but with coefficients $B_1$--$B_{19}$ in (\ref{aQ}) given as functions of the Lagrangian coefficients $A_1$--$A_9$ and the mass ratios $\chi_1$ and $\chi_2$.  Setting these coefficients equal to the values for $B_1$--$B_{19}$ given (\ref{Bs}) gives a system of 19 equations for the 9 unknowns $A_1$--$A_9$, which has the unique solution
\bea
A_1&=&\frac{3\chi_1}{4}(3+\eta),\quad
A_2=\frac{15\eta\chi_1}{4},
\nnm\\
A_3&=&-\frac{3\chi_1}{2}(1+3\chi_1),\quad
A_4=\frac{3\chi_1^2}{2},
\nnm\\
A_5&=&-\frac{3\chi_1^2}{2}(3+\chi_2),\quad
A_6=-\frac{3\eta}{2},
\nnm\\
A_7&=&-\frac{3\eta}{4},\quad
A_8=\frac{\chi_1^2}{2},\quad  \label{As}
A_9=\chi_1.
\eea
\ese

Thus, the action principle (\ref{pnL}) reproduces the 1PN CoM-frame equation of motion (\ref{aCoM}) for the relative position $z^i$---if we also make use of the evolution equations (\ref{Sdot}) and (\ref{E2intdot}) for the spin $S^i$ and internal energy $E^{\rm{int}}$ of body 2.  In the next section, we discuss an action principle that leads to a closed set of evolution equations for the binary system in the adiabatic approximation.

\section{Internal dynamics in the adiabatic approximation}\label{adiabatic}

\subsection{Euler-Lagrange equation for $Q^{ab}$}\label{ELQ}

We have just seen that the CoM-frame orbital EoM (\ref{aCoM}), the evolution equation for the binary's 1PN-accurate relative position $z^i(t)=z_2^i(t)-z_1^i(t)$, can be derived from the action principle (\ref{pnL}).  We now seek to extend this action principle to incorporate the internal dynamics of the deformable body 2 in the case where the quadrupole moment is adiabatically induced by the tidal field.

In Sec.~\ref{nadiabatic}, we saw how the adiabatic evolution of the quadrupole can encoded in a Newtonian action principle; varying the action (\ref{nLfinal}) with respect to the quadrupole $Q^{ij}$ gives
\bea\label{nQad}
Q^{ij}&=&\lambda \n G_2^{ij}+O(c^{-2})
\nnm\\
&&=\lambda\frac{3M_1}{r^3}n^{<ij>}+O(c^2),
\eea
for the Newtonian-order quadrupole (cf. (\ref{nQlambda})).  We also saw that, with the quadrupole given by (\ref{nQad}), the spin evolution equation becomes $\dot S^i=O(c^{-2})$ (cf.~Eq.~\ref{nadspin}), so that body 2 experiences no tidal torques.  For this reason, in the adiabatic case (unlike in the general case), we can specialize our analysis to the case of zero spin without generating inconsistencies, which we will do for the remainder of this section.

We have found that the simple Newtonian Lagrangian (\ref{nLfinal}) can be extended to govern the 1PN-accurate adiabatic evolution of $Q^{ij}$ in a relatively straightforward manner.  We consider the following Lagrangian:
\bse\label{pnadL}
\bea
\mc L&=&\mc L_{\rm{orb}}+\mc L^{\rm{int}}_2+O(c^{-4}),
\nnm\\
\mc L_{\rm{orb}}&=&\mc L_M+U^{ab}Q^{ab}+V^{ab}\dot Q^{ab}+WE^{\rm{int}},\label{Lorb}
\\
\mc L_2^{\rm{int}}&=&-\frac{1}{4\lambda}Q^{ab}Q^{ab},\label{pnLint}
\eea
\ese
with $\mc L_M(z,v)$ given by (\ref{LM}) and with $U^{ab}(z,v)$, $V^{ab}(z,v)$, and $W(z,v)$ being the coefficients appearing in $\mc L_Q$ (\ref{LQ}).  We have postulated (motivated by symmetry considerations) that the internal Lagrangian $\mc L_2^{\rm{int}}$ can still be taken as a simple quadratic in $Q^{ab}$ (\ref{pnLint}), generalizing the Newtonian Lagrangian (\ref{nLfinal}) only by using the 1PN-accurate value of $Q^{ab}$ in place of its Newtonian value.  (It is consistent to use here the fully relativistic value for the tidal deformability $\lambda$.)  The internal energy $E^{\rm{int}}$ appearing in $\mc L_{\rm{orb}}$ (\ref{Lorb}), which is needed only to Newtonian order, will still be given by
\be\label{adEint}
E^{\rm{int}}=\frac{1}{4\lambda}Q^{ab}Q^{ab}+O(c^{-2}),
\ee
up to a constant, as in (\ref{nadE}).  To avoid explicitly introducing an additional constant contribution to the internal energy, we can absorb any such contribution into the constant `Newtonian' mass monopole $\n M_2$ [cf.~Eq.~(\ref{M2split})].

Treating the relative position $z^i(t)$ and the quadrupole $Q^{ab}(t)$ as the independent dynamical variables in the Lagrangian (\ref{pnadL}), we still recover the orbital EoM (\ref{aCoM}) from the Euler-Lagrange equation for $z^i$ (with $S^a=0$), and that for $Q^{ab}$,
\be
\left(\frac{\doe}{\doe Q^{ab}}+\frac{d}{dt}\frac{\doe}{\doe\dot Q^{ab}}\right)\mc L=0,
\ee
gives
\be
Q^{ab}=2\lambda(1+W)(-U^{ab}+\dot V^{ab})+O(c^{-4}).
\ee
Using the coefficients $U^{ab}$, $V^{ab}$, and $W$ from the Lagrangian (\ref{Lorb},\ref{pnL}), we find that
\bse\label{pnQG}
\be\label{pnQ}
Q_{ab}=\lambda G_2^{ab}+O(c^{-4})
\ee
where $G_2^{ab}$ is the body-frame gravito-electric tidal moment defined in Sec.~\ref{body} and calculated in Appendix \ref{atm}:
\bea\label{pnG2}
G_2^{ab}&=&\frac{3\chi_1 M}{r^3}n^{<ab>}
\nnm\\
&&+\frac{1}{c^2}\frac{3\chi_1 M}{r^3}
\bigg[\left(2v^2-\frac{5\chi_2^2}{2}\dot{r}^2-\frac{5+\chi_1}{2}\frac{M}{r}\right)n_{<ab>}
\nnm\\
&&+v_{<ab>}-(3-\chi_2^2)\dot{r}n_{<a}v_{b>}\bigg]+O(c^{-4}),
\eea
\ese
The relation (\ref{pnQ}) between the body-frame tidal moment $G_2^{ab}$ and the adiabatically induced quadrupole $Q^{ab}$, derived here from the Lagrangian (\ref{pnadL}), is the same relation typically used to define the adiabatic approximation (e.g.~in \cite{DN2}).  The explicit expression for the tidal moment in (\ref{pnG2}) matches those given in Refs.~\cite{DSX3,DSX4,Poisson} when the latter are specialized to the CoM frame via (\ref{comvs}).

\subsection{Reduced Lagrangian, equations of motion, and conserved energy}\label{finalsec}

By substituting the solution (\ref{pnQG}) for the quadrupole into the Lagrangian (\ref{pnadL}), we find a reduced Lagrangian for the orbital dynamics involving only the CoM-frame orbital separation $z^i(t)$:
\bea\label{Lad}
\mc L[z^i]&=&\frac{\mu v^2}{2}+\frac{\mu M}{r}\left(1+\frac{\Lambda}{r^5}\right)
\nnm\\
&&+\frac{\mu}{c^2}\bigg\{\theta_0v^4
+\frac{M}{r}\bigg[v^2\left(\theta_1+\xi_1\frac{\Lambda}{r^5}\right)
\nnm\\
&&+\dot{r}^2\left(\theta_2+\xi_2\frac{\Lambda}{r^5}\right)
+\frac{M}{r}\left(\theta_3+\xi_3\frac{\Lambda}{r^5}\right)\bigg]\bigg\}\phantom{yo}
\eea
with
\be\label{Lambda}
\Lambda=\frac{3\chi_1}{2\chi_2}\lambda,
\ee
and with the dimensionless coefficients
\bea\label{Ladcs}
\theta_0&=&(1-3\eta)/8,
\nnm\\
\theta_1&=&(3+\eta)/2,
\nnm\\
\theta_2&=&\eta/2,
\nnm\\
\theta_3&=&-1/2,
\nnm\\
\xi_1&=&(\chi_1/2)(5+\chi_2),
\nnm\\ 
\xi_2&=&-3(1-6\chi_2+\chi_2^2),
\nnm\\
\xi_3&=&-7+5\chi_2.
\eea

While this form for the Lagrangian has been derived in harmonic gauge, we note that (some) other gauge choices lead to a Lagrangian with the same terms as in (\ref{Lad}) but with different values of the $\theta$ and $\xi$ coefficients.  In particular, the Lagrangian derived by BDF \cite{BDF} (when specialized to 1PN accuracy and to the center-of-mass frame) has this form, as does that obtained (via a Legendre transformation) from the EOB Hamiltonian including 1PN tidal effects proposed by Damour and Nagar \cite{DN2} (except that their work originally did not provide a value for the coefficient $\xi_2$).  In Appendix \ref{dam}, we derive the canonical transformation relating the EOB Hamiltonian to the harmonic-gauge Hamiltonian, which fixes the value of $\xi_2$ in the EOB Hamiltonian, and we present a separate transformation relating our results to those of BDF.  We thus demonstrate the complete equivalence of all of these results at 1PN order.

The orbital EoM resulting from the Lagrangian (\ref{Lad}) is given by
\bea\label{ada}
a^i&=&-\frac{Mn^i}{r}\left(1+\frac{6\Lambda}{r^5}\right)
\\
&&+\frac{M}{c^2r^2}\bigg[
v^2n^i\left(\phi_1+\zeta_1\frac{\Lambda}{r^5}\right)
+\dot{r}^2n^i\left(\phi_2+\zeta_2\frac{\Lambda}{r^5}\right)
\nnm\\ \nnm
&&+\frac{M}{r}n^i\left(\phi_3+\zeta_3\frac{\Lambda}{r^5}\right)
+\dot{r}v^i\left(\phi_4+\zeta_4\frac{\Lambda}{r^5}\right)\bigg],
\eea
with coefficients
\bea
\phi_1&=&4\theta_0-\theta_1-2\theta_2,
\nnm\\
\phi_2&=&3\theta_2,
\nnm\\
\phi_3&=&2(\theta_1+\theta_2-\theta_3),
\nnm\\
\phi_4&=&2(4\theta_0+\theta_1),
\nnm\\
\zeta_1&=&2(12\theta_0-3\xi_1-\xi_2),
\nnm\\
\zeta_2&=&8\xi_2,
\nnm\\
\zeta_3&=&12\theta_1+12\theta_2+2\xi_1+2\xi_2-7\xi_3,
\nnm\\
\zeta_4&=&12(4\theta_0+\xi_1),
\eea
for general values of the Lagrangian coefficients, and with
\bea
\phi_1&=&-1-3\eta,
\nnm\\
\phi_2&=&3\eta/2,
\nnm\\
\phi_3&=&2(2+\eta),
\nnm\\
\phi_4&=&2(2-\eta),
\nnm\\
\zeta_1&=&-3(2-\chi_2)(1+6\chi_2),
\nnm\\
\zeta_2&=&24(1-6\chi_2+\chi_2^2),
\nnm\\
\zeta_3&=&66+9\chi_2-19\chi_2^2,
\nnm\\
\zeta_4&=&6(2-\chi_2)(3-2\chi_2),
\eea
in harmonic gauge.

Finally, from the Lagrangian (\ref{Lad}), we can construct the conserved energy,
\bea\label{Ead}
E&=&v^i\doe\mc L/\doe v^i-\mc L
\nnm\\
&=&\frac{\mu v^2}{2}-\frac{\mu M}{r}\left(1+\frac{\Lambda}{r^5}\right)
\nnm\\
&&+\frac{\mu}{c^2}\bigg\{3\theta_0v^4
+\frac{M}{r}\bigg[v^2\left(\theta_1+\xi_1\frac{\Lambda}{r^5}\right)
\nnm\\
&&+\dot{r}^2\left(\theta_2+\xi_2\frac{\Lambda}{r^5}\right)
-\frac{M}{r}\left(\theta_3+\xi_3\frac{\Lambda}{r^5}\right)\bigg]\bigg\},\phantom{yo}
\eea
which is a constant of motion of the orbital EoM (\ref{ada}). 

As an application of these results, we can compute the gauge-invariant energy-frequency relationship for circular orbits.  Using the relations $\dot{r}=0$, $v^2=r^2\omega^2$, and $a^i=-r\omega^2n^i$ for a circular orbit, the orbital EoM (\ref{ada}) can be solved perturbatively, working to linear order both in the post-Newtonian parameter $1/c^2$ and in the tidal deformability parameter $\Lambda$ (\ref{Lambda}), to find the radius $r$ as a function of the orbital frequency $\omega$.  Combining this result with a similar treatment of the energy (\ref{Ead}), we can eliminate $r$ to find $E(\omega)$:
\bea
E(\omega)&=&\mu(M\omega)^{2/3}\bigg[-\frac{1}{2}+\frac{3\Lambda\omega^{10/3}}{M^{5/3}}
\nnm\\
&&\phantom{\mu(M\omega)^{2/3}\bigg[}+f_M\frac{(M\omega)^{2/3}}{c^2}+f_Q\frac{\Lambda\omega^4}{Mc^2}\bigg],
\nnm\\
f_M&=&\frac{1}{3}(\theta_0+\theta_1+\theta_3)=\frac{9+\eta}{24},
\nnm\\
f_Q&=&\frac{11}{3}(8\theta_0+2\theta_1-4\theta_3+\xi_1+\xi_3)
\nnm\\
&=&\frac{11}{6}(3+2\chi_2+3\chi_2^2).
\eea
While the $\theta$ and $\xi$ coefficients may take different values in different gauges, their combinations appearing here must be gauge-invariant.  As $E(\omega)$ is independent of $\xi_2$, we can check this result against those obtained from the EOB Hamiltonian of Damour and Nagar \cite{DN2} (see Appendix \ref{dam}), and we find that they agree.

\section{Conclusion}

We have derived the first-post-Newtonian orbital equations of motion for binary systems of bodies with spins and mass quadrupole moments, at linear order in the spin and quadrupole, and shown that they conserve the total linear momentum of the binary.  After specializing these results to the binary's center-of-mass-energy frame, we have found an action principle from which the orbital equations of motion can be derived.  Finally, we considered the case in which the quadrupole moment is adiabatically induced by the tidal field, giving a simplified Lagrangian and equation of motion for this case, as well as the conserved energy function and the energy-frequency relationship for circular orbits.  These results are useful for the calculation of tidal effects in the gravitational wave signals from inspiralling neutron star binaries.

\begin{acknowledgments}

The authors would like to thank Tanja Hinderer and Etienne Racine for many helpful discussions and comments.  This work was supported by the New York State NASA Space Grant Consortium and by NSF Grant PHY-0757735.

\end{acknowledgments}

\appendix

\section{Hamiltonian for the adiabatic orbital dynamics, canonical transformations, and comparison with DN and BDF}\label{dam}

Our aim here is to demonstrate the equivalence of the results for the orbital-tidal conservative dynamics given by Damour and Nagar \cite{DN2}, Bini, Damour and Faye \cite{BDF}, and the present work.  We begin with a discussion of Hamiltonians and canonical transformations, then relate our results to the EOB Hamiltonian given by DN (also revisited and completed by BDF), and finish by relating the Lagrangian given by BDF to ours.

As in Sec.~\ref{finalsec}, we consider a Lagrangian for the CoM-frame orbital separation $z^i=z_2^i-z_1^i$ of the form
\bea\label{BLad}
\hat{\mc{L}}&=&\frac{v^2}{2}+\frac{1}{r}\left(1+\frac{\Lambda}{r^5}\right)
\nnm\\
&&+\frac{1}{c^2}\bigg[\theta_0v^4
+\frac{v^2}{r}\left(\theta_1+\xi_1\frac{\Lambda}{r^5}\right)
+\frac{\dot{r}^2}{r}\left(\theta_2+\xi_2\frac{\Lambda}{r^5}\right)
\nnm\\
&&
+\frac{1}{r^2}\left(\theta_3+\xi_3\frac{\Lambda}{r^5}\right)\bigg]+O(c^{-4}),\phantom{yo}
\eea
in units that set $M=1$, and with the Lagrangian having been rescaled by the symmetric mass ratio, $\hat{\mc{L}}=\mc L/\eta$.  We have shown that the harmonic-gauge values of the $\theta$ and $\xi$ coefficients are given by Eq.~(\ref{Ladcs}).

The (rescaled) momentum canonically conjugate to $z^i$ is
\bea
p^i= \frac{\doe \hat{\mc{L}}}{\doe v^i}&=&v^i+\frac{1}{c^2}\bigg[4\theta_0v^2v^i+\frac{2v^i}{r}\left(\theta_1+\xi_1\frac{\Lambda}{r^5}\right)
\nnm\\
&&+\frac{2\dot r n^i}{r}\left(\theta_2+\xi_2\frac{\Lambda}{r^5}\right)\bigg]+O(c^{-4})
\nnm\\
&\equiv & v^i+\frac{1}{c^2}\delta p^i(\bm{z},\bm{v})+O(c^{-4}).
\eea
From a Legendre transformation of the Lagrangian $\hat{\mc{L}}(\bm{z},\bm{v})$, we can construct the Hamiltonian:
\bea
\hat{H}(\bm{z},\bm p)&=&p^jv^j-\hat{\mc{L}}(\bm{z},\bm{v})
\nnm\\
&=&p^2-\!\frac{1}{c^2}p^j\delta p^j-\hat{\mc{L}}(\bm{z},\bm{p})+\!\frac{1}{c^2}\frac{\doe \hat{\mc{L}}}{\doe v^j}\delta p^j\!+\!O(c^{-4})
\nnm\\
&=&p^2-\hat{\mc{L}}(\bm{z},\bm{p})+O(c^{-4}),
\eea
having used $\delta p^i(\bm{z},\bm{v})=\delta p^i(\bm{z},\bm{p})+ O(c^{-4})$.  This gives
\bea\label{H}
\hat{H}&=&\frac{v^2}{2}-\frac{1}{r}\left(1+\frac{\Lambda}{r^5}\right)
\nnm\\
&&-\frac{1}{c^2}\bigg[\theta_0v^4
+\frac{v^2}{r}\left(\theta_1+\xi_1\frac{\Lambda}{r^5}\right)
+\frac{\dot{r}^2}{r}\left(\theta_2+\xi_2\frac{\Lambda}{r^5}\right)
\nnm\\
&&
+\frac{1}{r^2}\left(\theta_3+\xi_3\frac{\Lambda}{r^5}\right)\bigg]+O(c^{-4}),\phantom{yo}
\eea
Note that the effect of the Legendre transformation has been simply to flip the sign of all terms except the first.

In Ref.~\cite{DN2}, Damour and Nagar consider an EOB Hamiltonian of the form
\be
\hat{H}_{\rm{EOB}}=\frac{1}{\eta}\left[1+2\eta (\hat{H}_{\rm{eff}}-1)\right]^{1/2},
\ee
with
\be
\hat{H}_{\rm{eff}}=\left[\frac{A}{B}\frac{p_r^2}{c^2}+A\left(1+\frac{p_\phi^2}{c^2r^2}\right)\right]^{1/2},
\ee
Here, $p_r$ and $p_\phi$ are the momenta conjugate to the polar coordinates $(r,\phi)$ in the plane of motion, related to the Cartesian momenta used above by $p_r=\bm n \cdot \bm p$ and $p_\phi^2/r^2=p^2-p_r^2$.  The functions $A(r)$ and $B(r)$ are coefficients in the EOB effective metric \cite{DN0}; $A(r)$ completely encodes the energetics of circular orbits, while $B(r)$ has effects only when $p_r\ne0$.  Damour and Nagar proposed to incorporate Newtonian and 1PN tidal effects into this EOB Hamiltonian by adding tidal terms to the radial potential $A$:
\be
A(r)=1-\frac{2}{c^2r}-\frac{2\Lambda}{c^2r^6}\left(1+\frac{\alpha_1}{c^2 r}\right)+O(c^{-6}),
\ee
for the quadrupole $l=2$ case.  They have computed the 1PN tidal coefficient to be
\be
\alpha_1=\frac{5}{2}\chi_2.
\ee
While they did not propose to modify the potential $B(r)$ from its point-particle value of $B=1+2/c^2r+O(c^{-4})$, we find that such a modification,
\be
B(r)=1+\frac{2}{c^2r}\left(1+\beta_0 \frac{\Lambda}{r^5}\right)+O(c^{-4})
\ee
for some coefficient $\beta_0$, is necessary to match our results.  Expanding the EOB Hamiltonian with these values for the potentials, we find a Hamiltonian of the form (\ref{H}) with coefficients
\bea\label{damcs}
\bar\theta_0&=&(1+\eta)/8,
\nnm\\
\bar\theta_1&=&(1-\eta)/2,
\nnm\\
\bar\theta_2&=&1,
\nnm\\
\bar\theta_3&=&(1+\eta)/2,
\nnm\\
\bar\xi_1&=&(1-\eta)/2,
\nnm\\ 
\bar\xi_2&=&\beta_0,
\nnm\\
\bar\xi_3&=&1+\eta+\alpha_1,
\eea
instead of the harmonic-gauge coefficients in Eq.~(\ref{Ladcs}).

Without tidal effects, the EOB Hamiltonian and the harmonic-gauge Hamiltonian (which coincides with the ADM Hamiltonian at 1PN order) are known to be related by a canonical transformation \cite{DN0}.  Considering a 1PN-order canonical transformation with generating function $G$,
\be
z^i\to z^i+\frac{1}{c^2}\frac{\doe}{\doe p^i}G(\bm{z},\bm{p}),\quad p^i\to p^i-\frac{1}{c^2}\frac{\doe}{\doe z^i}G(\bm{z},\bm{p}),
\ee
the Hamiltonian changes by
\be
\hat H\to \hat H+\frac{1}{c^2}\{\hat H,G\}+O(c^{-4}),
\ee
with only the Newtonian part of the Hamiltonian contributing in the Poisson bracket.  We find that the most general generating function $G$ that preserves the form of the Hamiltonian including tidal effects (\ref{H}), changing its coefficients but adding no new terms, is of the form
\be\label{cc}
G=(\bm z \cdot \bm p)\left(\gamma_1 p^2 + \gamma_2 \frac{1}{r} + \gamma_3 \frac{\Lambda}{r^6} \right)
\ee
with arbitrary constant $\gamma$ coefficients.  The changes in the Hamiltonian coefficients induced by the canonical transformation are
\bea\label{deltas}
\Delta\theta_0&=&-\gamma_1,
\nnm\\
\Delta\theta_1&=&\gamma_1-\gamma_2,
\nnm\\
\Delta\theta_2&=&2\gamma_1+\gamma_2,
\nnm\\
\Delta\theta_3&=&\gamma_2,
\nnm\\
\Delta\xi_1&=&6\gamma_1-\gamma_3,
\nnm\\ 
\Delta\xi_2&=&12\gamma_1+6\gamma_3,
\nnm\\
\Delta\xi_3&=&6\gamma_2+\gamma_3,
\eea

If we set the EOB Hamiltonian coefficients (\ref{damcs}) equal to the harmonic Hamiltonian coefficients (\ref{Ladcs}) plus the transformation parameters (\ref{deltas}), we find that this (redundant) system of equations has a unique solution.  The coefficients in the canonical transformation (\ref{cc}) must be
\bea
\gamma_1&=&-\eta/2,
\nnm\\
\gamma_2&=&(2+\eta)/2,
\nnm\\
\gamma_3&=&2-9\chi_2/2+2\chi_2^2,
\eea
with the values for $\gamma_1$ and $\gamma_2$ matching those computed in Ref.~\cite{DN0}, and the parameters in the EOB potentials must be
\bea
\alpha_1&=&5\chi_2/2,
\nnm\\
\beta_0&=&3(3-5\eta).
\eea
The value for $\alpha_1$ matches that given by Damour and Nagar, and the value for $\beta_0$ can be used to extend the range of validity of their EOB Hamiltonian to non-circular orbits.

We turn now to the more recent work of BDF \cite{BDF}.  They have derived a Lagrangian for the orbital-tidal conservative dynamics including terms up to 2PN order.  Here, we restrict attention to the 0PN and 1PN terms, specialize their results to the center-of-mass frame, and translate their notation into ours.  Their Lagrangian [whose tidal part can be found from their Eqs.~(2.12), (4.3), (4.4) and (4.10)] is of the form (\ref{BLad}) with the same point-mass ($\theta$) coefficients coefficients, but with tidal ($\xi$) coefficients 
\bea\label{BDFxis}
\tilde\xi_1 &=& - \chi_1^2/2 + 3
\nnm\\
\tilde\xi_2 &=& -3 -3\chi_2^2
\nnm\\
\tilde\xi_3 &=& -7 +2\chi_2.
\eea
The corresponding Hamiltonian is again given by Eq.~(\ref{H}).  One can then verify that a canonical transformation of the form (\ref{cc}) with
\be
\gamma_1=\gamma_2=0,\quad\gamma_3=3\chi_2,
\ee
using (\ref{deltas}), will transform their $\xi$ coefficients from Eq.~(\ref{BDFxis}) into ours from Eq.~(\ref{Ladcs}).  This demonstrates the complete equivalence of our respective results.

\section{Moment transformations and translational equations of motion}\label{atm}

We present here the formulae that relate the body-frame multipole and tidal moments and the global-frame multipole and tidal moments.  Such formulae were first derived by Damour, Soffel, and XU \cite{DSX1,DSX2,DSX3,DSX4} (see, in particular, Sec.~VD of Ref.~\cite{DSX2}).  They are derived by requiring the equivalence of the body- and global-frame metrics in the body's buffer region $\mc{B}^A$.  More specifically, one substitutes the expansions of the body-frame potentials (\ref{potA}) and the global-frame potentials (\ref{potgtidal}), along with the coordinate transformation (\ref{xt}), into the tensor transformation law for the metric (\ref{gtransf}), using (\ref{metric}) to express the metrics in terms of the potentials in both coordinate systems.  Matching coefficients of the resultant multipole expansions gives the moment transformation formulae.  Having used the body-frame gauge conditions (\ref{bfgauge}) to eliminate the worldline-data functions $\alpha_A$ and $\beta_A$, one finds that the transformation formulae involve only the various moments and the CoM worldlines $z_A^i$.  A more detailed account of the procedure is given in RF \cite{RF} Sec.~V.

After presenting the various moment transformation formulae in Secs.~\ref{bfmult_gfmult}, \ref{gfmult_gftidal}, and \ref{gftidal_bftidal}, we show in Sec.~(\ref{team}) how to use them to arrive at the translational equations of motion for an $N$-body system.

Note that \emph{all} the quantities appearing below (moments and worldlines) are treated as functions of the global time coordinate $t$ (with the argument suppressed).  Any quantities originally defined as functions of the body-frame time coordinate $s_A$ (like the body-frame moments or the worldlines $z_A^i$) can be converted to functions of $t$ by using the coordinate transformation (\ref{xt}) at $y_A^i=0$, which gives
\be\label{sA0}
s_A=s_A^0(t)=t-c^{-2}\alpha_A(s_A)\big|_{s_A=t}+O(c^{-4}).
\ee
This change of variables does affect the forms of any equations presented here (or elsewhere in the paper).  It is important to note that our use of $f(s_A)$ and $f(t)$ to denote the same physical quantity differs from the convention of (e.g.) RF.  There, symbols are used to denote functions of a particular variable, not physical quantities, and $f(s_A)$ and $f(t)$ would be different physical quantities.  

\subsection{Body-frame multipole moments $\to$ global-frame multipole moments}\label{bfmult_gfmult}

First, via the metric transformation law, the global-frame multipole moments $M_\gA^L$ and $Z_\gA^{iL}$ [defined by (\ref{potg})] are given in terms of the body-frame moments $M_{A}^L$ and $S_{A}^{L}$ [defined by (\ref{potA})] and the CoM worldlines $z_A^i$, by
\begin{widetext}
\bea
M_\gA^L&=&M_A^L+\frac{1}{c^2}\bigg[
\left(\frac{3}{2}v_A^2-(l+1)G_\gA\right)M_A^L
-\frac{2l^2+5l-5}{(l+1)(2l+3)}v_A^j\dot M_A^{jL}\label{MgA}
\\  \nnm
&&-\frac{2l^3+7l^2+16l+7}{(l+1)(2l+3)}a_A^j M^{jL}_A - \frac{2l^2+17l-8}{2(2l+1)}v_A^{j<a_l}M_A^{L-1>j}+\frac{4l}{l+1}v_A^j\epsilon^{jk<a_l}S^{L-1>k}\bigg]+O(c^{-4}),
\\
Z^{iL}_\gA &=& \frac{4}{l+1}\dot{M}^{iL}_A + 4 v_A^i M^L_A
- \frac{4 (2l-1)}{2l+1} v_A^j M_A^{j<L-1} \delta^{a_l>i} - \frac{4l}{l+1}
\epsilon^{ji<a_l} S_A^{L-1>j}+O(c^{-2}),\label{ZgA}
\eea
\end{widetext}
Recall that $v_A^i=\dot z_A^i$ and $a_A^i=\ddot z_A^i$; the accelerations $a_A^i$ here (and anywhere they appear in 1PN-order terms) may be replaced here with their Newtonian values from (\ref{newteom2}).  The global-frame monopole tidal moments $G_\gA$ appearing here (and needed only with Newtonian accuracy here) can be expressed in terms of the body-frame multipole moments of bodies $B\ne A$ and the bodies' worldlines by using Eq.~(\ref{nGgArel}).

The specific (nonzero) instances of these formulae needed in the $M_1$-$M_2$-$S_2$-$Q_2$ system are as follows (with $M_2^{ij}\equiv Q^{ij}$ and $S_2^i\equiv S^i$):
\begin{widetext}
\bea
M_\go&=& M_1 + \frac{M_1}{c^2}\left(\frac{3}{2} v^2_1 - \frac{M_2}{r} - \frac{3}{2r^3}n^{jk}Q^{jk}\right)+O(c^{-4}),\label{MgAQ}
\\
M_\gt&=& M_2 + \frac{M_2}{c^2}\left(\frac{3}{2} v^2_2 - \frac{M_1}{r}\right)+O(c^{-4}),
\nnm\\
M_\gt^i&=& \frac{1}{c^2}\left(-\frac{1}{5} v_2^j \dot{Q}^{ij} + \frac{16M_1}{5r^2} n^j Q^{ij} +  2\epsilon^{ijk} v_2^j S^k\right)+O(c^{-4}),
\nnm \\ \nnm
M_\gt^{ij}&=&Q^{ij}+\frac{1}{c^2}\left[\left(\frac{3}{2}v_2^2-\frac{3M_1}{r}\right)Q^{ij}-\frac{17}{5}v_2^{k<i}Q^{j>k}\right]+O(c^{-4}),
\eea
\end{widetext}
and
\bea
Z_\go^i&=&4M_1 v_1^i+O(c^{-2}), \label{ZgAQ}
\\
Z_\gt^i&=&4M_2 v_2^i+O(c^{-2}),
\nnm\\
Z_\gt^{ij}&=&2\dot{Q}_2^{ij}-2\epsilon^{ijk}S_2^k+O(c^{-2}),
\nnm\\ \nnm
Z_\gt^{ijk}&=&4v_2^i Q_2^{jk} - \frac{12}{5} v_2^a Q_2^{a<j} \delta^{k>i}+O(c^{-2}),
\eea
Here, $r=|\bm z_2-\bm z_1|$ and $n^i=(z_2^i-z_1^i)/r$, as in (\ref{rel}).

\subsection{Global-frame multipole moments $\to$ global-frame tidal moments}\label{gfmult_gftidal}

Next, one can express the global-frame tidal moments $G_\gA^L(t)$ and $Y_\gA^{iL}(t)$ (def) a body $A$ in terms of the global-frame multipole moments $M_\gA^L$ and $Z_\gA^{iL}$ (def) of all the other bodies $B\ne A$ and the all the bodies' worldlines $z_C^i$.  For this purpose, rather than working directly with $G_\gA^L$, it is easier to work with the tensors $F_\gA^L$ defined by
\begin{widetext}
\bea
\Phi_{\g,\rm{ext}}&=&-\sum_{l=0}^\infty\frac{1}{l!}\left\{ G_\gA^L(x-z^A)^L+\frac{1}{2(2l+3)c^2}\doe_t^2\left[G_\gA^L(x-z_A)^{jjL}\right]\right\}+O(c^{-4})
\nnm\\ \label{defF}
&=&-\sum_{l=0}^\infty\frac{1}{l!}\left\{ F_\gA^L(x-z^A)^L+\frac{1}{2(2l+3)c^2}J_\gA^L(x-z_A)^{jjL}\right\}+O(c^{-4}),
\eea
\end{widetext}
cf.~(\ref{potgtidal}).  The tensor $J_\gA^L$ will not be needed in our calculations (and contains no extra information).  Note that $F_\gA^L$ and $G_\gA^L$ agree at Newtonian order,
\bdm
F_\gA^L=G_\gA^L+O(c^{-2}).
\edm
The Newtonian part of $G_\gA^L$ is also given in Eq.~(\ref{nGgArel}).

By equating the Eq.~(\ref{defF}) and the external part of Eq.~(\ref{potg}), one finds that $F_\gA^L$ is given by
\bea\label{FgA}
F_\gA^{L}&=&\sum_{B\ne A}\sum_{k=0}^\infty \frac{(-1)^k}{k!}\bigg[N_\gB^{K}\doe^{(A)}_{KL}\frac{1}{|\bm{z}_A-\bm{z}_B|}\phantom{yoyo}
\\ \nnm
&&+\frac{1}{2c^2}P_\gB^{K}\doe^{(A)}_{K<L>}|\bm{z}_A-\bm{z}_B|\bigg]+O(c^{-4}),
\eea
where the tensors $N_\gA^L$ and $P_\gA^L$ are given by
\bea
N_\gA^L&=&M_\gA^L+\frac{1}{(2l+3)c^2}\bigg[v_A^2M_A^L+2v_A^j\dot M_A^{jL}
\nnm \\ \nnm
&&+2lv_A^{j<a_l}M_A^{L-1>j}+a_A^jM_A^{jL}\bigg]+O(c^{-4}),
\\ \nnm
P^L_\gA &=& \ddot{M}^L_A + 2 l v_A^{<a_l}
\dot{M}_A^{L-1>}
+ l a_A^{<a_l} M_A^{L-1>}
\\ \nnm
&& + l(l-1) v_A^{<a_la_{l-1}} M_A^{L-2>}+O(c^{-2}).
\eea
Similarly, in the gravito-magnetic sector, one finds
\be\label{YgA}
Y_\gA^{iL}=\sum_{B\ne A}\sum_{k=0}^\infty \frac{(-1)^k}{k!}Z_\gB^{iK}\doe^{(A)}_{KL}\frac{1}{|\bm{z}_A-\bm{z}_B|}.
\ee

For the $M_1$-$M_2$-$S_2$-$Q_2$ system, using the non-zero global-frame multipole moments from the last subsection, one must compute the $l=0,1,2,3$ (resp.~$l=0,1$) cases of (\ref{FgA}) and the $l=0,1,2$ (resp.~$l=0$) cases of (\ref{YgA}) for $A=2$, $B=1$ (resp.~$A=1$, $B=2$).  The derivatives appearing here can be easily expressed in terms of $r=|\bm z_2-\bm z_1|$ and $n^i=(z_2^i-z_1^i)/r$ via the identities
\bea
\doe^{(1)}_{L}\frac{1}{r}&=&(-1)^l\doe^{(2)}_{L}\frac{1}{r}=\frac{1}{(2l+1)!!}\frac{n^{<L>}}{r^{l+1}},
\nnm\\ \nnm
\doe_L r&=&\frac{r^2}{2l-1}\doe_L\frac{1}{r}+\frac{l(l-1)}{2l-1}\delta_{(a_la_{l-1}}\doe_{L-2)}\frac{1}{r}.
\\ \nnm &&
\eea

\subsection{Global-frame tidal moments $\to$ body-frame tidal moments}\label{gftidal_bftidal}

Finally, the metric transformation law gives the body-frame tidal moments $G_A^L$ and $H_A^L$ (defined by (\ref{potA})) in terms of the global-frame tidal moments $F_\gA^L$ and $Y_\gA^L$ (defined by (\ref{defF}),(\ref{potgtidal})).  First, the gravito-magnetic tidal moments can be found from
\bdm
H_A^L=Y_A^{jk<L-1}\epsilon^{a_l>jk},
\edm
with the tensors $Y_A^{iL}$ given by
\be
Y_A^{iL}=(\delta_{jK}^{iL}-\delta_{jK}^{<iL>})(Y_\gA^{jK}-4v_A^j \n G_\gA^{K} - l! \Lambda_\zeta^{jK}).
\ee
Here, $\delta_{jK}^{iL}$ is the multi-Kronecker delta ($\delta_{jK}^{iL}T^{jK}=T^{iL}$), and $\delta_{jK}^{<iL>}$ is the STF projector ($\delta_{jK}^{<iL>}T^{jK}=T^{<iL>}$), and the nonzero 'inertial moments' $\Lambda_\zeta^{iL}$ are
\bea
\Lambda_\zeta^i&=&- 2 G_\gA v_A^i,
\nnm \\
\Lambda_\zeta^{ij}&=&-\frac{3}{2}v_A^{[i} a_A^{j]}-2v_A^{<i} a_A^{j>} -\frac{4}{3} \dot G_\gA \delta^{ij},
\nnm \\ \nnm
\Lambda_\zeta^{ijk}&=&-\frac{6}{5}\delta^{i<j} \dot a_A^{k>}.
\eea
Then, the gravito-electric tidal moments are given by
\begin{widetext}
\bea
\pn G_A^L&=& F_\gA^L+ - l!\Lambda_\Phi^L + \frac{1}{c^2}\bigg[\dot Y_\gA^{<L>}-v_A^jY_\gA^{jL}
+ ( 2 v_A^2 - l G_\gA ) G_\gA^L - (l/2) v_A^{j<a_l}G_\gA^{L-1>j}
\nnm\\
&&+(l-4)v_A^{<a_l}\dot G_\gA^{L-1>}-(l^2-l+4)a_A^{<a_l}G_\gA^{L-1>}-(l-1)!\dot \Lambda_\zeta^{<L>}\bigg]+O(c^{-4}),
\eea
\end{widetext}
for $l\ge 1$ and by $G_A=0$ for $l=0$ (cf.~(\ref{zeropotential})).  The non-zero inertial moments $\Lambda_\Phi^L$ needed here are
\bea
\Lambda_\Phi^i&=&a_A^i+\frac{1}{c^2}\bigg[\left(v_A^2+G_\gA\right)a_A^i
+\frac{1}{2}v_A^{ij}a_A^j+2\dot G_\gA v_A^i\bigg],
\nnm \\ \nnm
\Lambda_\Phi^{ij}&=&\frac{1}{c^2}\left(-\frac{1}{2}a_A^{<ij>}+v_A^{<i}\dot a_A^{j>}\right).
\eea

\subsection{Translational equations of motion}\label{team}

The results of Secs.~\ref{bfmult_gfmult}, \ref{gfmult_gftidal}, and \ref{gftidal_bftidal} allow one to express the body-frame tidal moments $G_A^L$ and $H_A^L$ for a given body $A$ in terms of the body-frame multipole moments $M_B^L$ and $S_B^L$ and CoM worldlines $z_B^i$ of all bodies $B$ in an $N$-body system (eliminating all reference to the global frame moments).  With this done, one can find the body's translational equation of motion, written only in terms of the $M_B^L$, $S_B^L$, and $z_B^i$, by using the law of motion (\ref{lom_mi}) for the body-frame mass dipole $M_A^i$.

As the body-frame gauge condition (\ref{masscenter}) requires $M_A^i=0$ (fixing the body's center of mass-energy to the body-frame origin), one proceeds by setting the right-hand side of (\ref{lom_mi}) to zero.  This yields an expression for the acceleration $a_A^i=\ddot z_A^i$ of the body's 1PN-accurate global-frame CoM worldline $z_A^i(t)$.  To see this more clearly, we can explicitly evaluate the $l=0$ case of the first term on the RHS of (\ref{lom_mi}) using the moment transformation formulae presented above.  The result is
\begin{widetext}
\bea
M_A a_A^i&=&M_A F_\gA+\sum_{l=2}^\infty \frac{1}{l!} M_A^L G_A^L + \frac{1}{c^2}g_A^i
\\ \nnm
&&+\frac{1}{c^2}\bigg[\dot Y_\gA^i-v_A^jY_\gA^{ji} + (2v_A^2-G_\gA)G_\gA^i-\frac{1}{2}v_A^{ij}G_\gA^j
- (v_A^2+3 G_\gA) a_A^i -\frac{1}{2}v_A^{ij}a_A^j - 3 \dot G_\gA v_A^i \bigg]+O(c^{-4}),
\eea
\end{widetext}
where $g_A^i$ represents the terms on the RHS of Eq.~(\ref{lom_mi}) except for the first.

\section{Derivation of system mass multipole moment formulae}\label{sysmmm}

We derive here Eq.~(\ref{MLsys}), which gives the mass multipole moments $M_{\rm{sys}}^L$ of an $N$-body system in terms of its global frame multipole moments $M_\gA^L$ and $Z_\gA^L$.  This parallels the derivation given in Sec.~(\ref{genform}) for the moments $Z_{\rm{sys}}^{iL}$.  We begin by using the Taylor series (\ref{taylor}) to rewrite the global-frame scalar potential $\Phi_\g$ as given in Eq.~(\ref{phigC}) in the form
\begin{widetext}
\bea
\Phi_{\rm{g}}&=&-\sum_A\sum_{l,k=0}^\infty \frac{(-1)^{l+k}}{l!k!}\left[M_\gA^L z_A^K\doe_{LK}\frac{1}{|\bm x|}
+\frac{1}{2c^2}\doe_t^2\left(M_\gA^L z_A^K\doe_{LK}|\bm x|\right)\right]
\nnm\\
&=&-\sum_A\sum_{p=0}^\infty\sum_{k=0}^p \frac{(-1)^{p}}{p!}\frac{p!}{k!(p-k)!}\left[M_\gA^{<P-K} z_A^{K>}\doe_{P}\frac{1}{|\bm x|}
+\frac{1}{2c^2}\doe_t^2\left( M_\gA^{(P-K} z_A^{K)}\right)\doe_{P}|\bm x|\right].
\eea
\end{widetext}
To bring this into the form (\ref{phigsys}), which gives $\Phi_\g$ in terms of the system moments, we must decompose $\doe_P|\bm x|$ into its STF and trace parts.  Using the identity
\be
\doe_{ijL}|x|=\doe_{<ijL>}|x|+\frac{(l+1)(l+2)}{2l+3}\delta_{(ij}\doe_{L)}\frac{1}{|x|},
\ee
and making the index change $P\to ijL$, but only in the term resulting from the trace part of $\doe_P|x|$, we find
\begin{widetext}
\bea
\Phi_\g&=&-\sum_A\sum_{p=0}^\infty\sum_{k=0}^p \frac{(-1)^{p}}{p!}\frac{p!}{k!(p-k)!}\left[M_\gA^{<P-K} z_A^{K>}\doe_{P}\frac{1}{|\bm x|}
+\frac{1}{2c^2}\doe_t^2\left( M_\gA^{<P-K} z_A^{K>}\right)\doe_{<P>}|x|\right]
\nnm\\
&&-\frac{1}{2c^2}\sum_A\sum_{l=0}^\infty\sum_{k=0}^{l+2}\frac{(-1)^{l}}{(l+2)!}\frac{(l+2)!}{k!(l+2-k)!}
\doe_t^2\left( M_\gA^{(ijL-K} z_A^{K)}\right)\frac{(l+1)(l+2)}{2l+3}\delta_{ij}\doe_{L}\frac{1}{|x|}\label{secondline}
\eea
\end{widetext}
Though the sum over $l$ should start at $l=-2$ after $P\to ijL$, the $l=-1,-2$ terms are killed by the $(l+1)(l+2)$ factor.  In the $\sum_{k=0}^{l+2}$ term, the $k$ indices $K$ are to be chosen from the $l+2$ indices $ijL$.  This term can be simplified by explicitly performing the symmetrization over all $l+2$ indices; using the fact that $ M_\gA^L$ and $\doe_L|\bm x|^{-1}$ are STF, we have
\begin{widetext}
\be
 M_\gA^{(ijL-K} z_A^{K)}\delta^{ij}\doe_{L}\frac{1}{|\bm x|}=\left[\frac{2k(l+2-k)}{(l+2)(l+1)} M_\gA^{j<L-(K-1)} z_A^{K-1>j}
+\frac{k(k-1)}{(l+2)(l+1)} M_\gA^{<L-(K-2)} z_A^{K-2>jj}\right]\doe_{L}\frac{1}{|\bm x|}
\ee
Using this identity, relabeling $K-1\to K$ in the first term and $K-2\to K$ in the second term, and adjusting summations appropriately, we find that the second line of (\ref{secondline}) can be written as
\be
-\frac{1}{2(2l+3)c^2}\sum_A\sum_{l=0}^\infty\sum_{k=0}^{l}\frac{(-1)^{l}}{l!}\frac{l!}{k!(l-k)!}
\doe_t^2\left(   
2 M_\gA^{j<L-K} z_A^{K>j}
+ M_\gA^{<L-K} z_A^{K>jj}
   \right)\doe_{L}\frac{1}{|x|}\label{replace}
\ee
Finally, we can compare (\ref{secondline}) (with the second line replaced by (\ref{replace})) to the expression for $\Phi_\g$ given in (\ref{phigsys}); we see that the system's 1PN-accurate mass multipoles must be given by
\bea
M_{\rm{sys}}^L&=&\sum_A\sum_{k=0}^l\frac{l!}{k!(l-k)!}\left[ M_\gA^{<L-K} z_A^{K>}
+\frac{1}{c^2}\frac{1}{2(2l+3)}\doe_t^2\left(2 M_\gA^{j<L-K} z_A^{K>j}+ M_\gA^{<L-K} z_A^{K>jj}\right)\right]
\nnm\\ \nnm
&&-\frac{1}{c^2}\frac{2l+1}{(l+1)(2l+3)}\dot{\mu}_{\rm{sys}}^L+O(c^{-4}),
\eea
as in Eq.~(\ref{MLsys}).
\end{widetext}

\bibliography{bt}

\end{document}